\documentclass[prc,aps,superscriptaddress
,unsortedaddress,groupedaddress,amsmath,amssymb,twocolumn]{revtex4}

\usepackage{amsmath}    % need for subequations
\usepackage{graphicx}   % need for figures
\usepackage{verbatim}   % useful for program listings
\usepackage{color}      % use if color is used in text
\usepackage{subfigure}  % use for side-by-side figures
\usepackage{hyperref}   % use for hypertext links, including those to external documents and URLs
\usepackage{aas_macros}

\def\msun{{\rm M}_\odot}

\def\drom{{\rm d}}

\begin{document}
%%%%%%%%%%%%%%%%%

%%%%%%%%%%%%%%%%%
\title{Neutron star radii and crusts: uncertainties and unified equations of state}
\author{M. Fortin}
\affiliation{N. Copernicus Astronomical Center, Polish Academy of Sciences, Bartycka 18, 00-716 Warszawa, Poland}
\author{C. Provid\^encia} 
\affiliation{CFisUC, Department of Physics, University of Coimbra, P-3004-516 Coimbra, Portugal}
\author{A. R. Raduta} 
\affiliation{IFIN-HH, Bucharest-Magurele, POB-MG6, Romania}
\author{F. Gulminelli} 
\affiliation{Normandie Univ., ENSICAEN, UNICAEN, CNRS/IN2P3, LPC Caen, 14000 Caen, 
France}
\author{J. L Zdunik} 
\author{P. Haensel}
\affiliation{N. Copernicus Astronomical Center, Polish Academy of Sciences, Bartycka 18, 00-716 Warszawa, Poland}
\author{M. Bejger}
\affiliation{N. Copernicus Astronomical Center, Polish Academy of Sciences, Bartycka 18, 00-716 Warszawa, Poland}

\date{\today}
%%%%%%%%%%%%%%%%%

%%%%%%%%%%%%%%%%%
\begin{abstract}
The uncertainties in neutron star radii and crust properties due to our limited knowledge of the equation of state are quantitatively analysed.
We first demonstrate the importance of a unified microscopic description 
for the different baryonic densities of the star.  If  the pressure functional
is obtained matching a crust and a core equation of
state  based on models with different 
properties at nuclear matter saturation, the uncertainties can be as large as $\sim$ 30\% for the crust
thickness and 4\% for the radius.   
Necessary conditions for causal and thermodynamically consistent matchings between the core and the crust are formulated and their consequences examined.
A large set of unified equations of state for purely nucleonic matter is obtained based on twenty four Skyrme interactions and nine relativistic mean-field
nuclear parametrizations. In addition, for relativistic models seventeen equations of state including a transition to hyperonic matter at high density are presented.
All these equations of state have in common
the property of describing a $2\;M_\odot$ star and of being causal within stable neutron stars. A span  of $\sim 3$ km  and $\sim 4$ km is obtained for the
radius of, respectively, $1.0\;M_\odot$ and $2.0\;M_\odot$ star. Applying a set of nine further constraints from experiment and ab-initio calculations the
uncertainty is reduced to $\sim$ 1 km and 2 km, respectively. 
These residual uncertainties
reflect lack of constraints at large densities and insufficient
information on the density dependence of the equation of state near the nuclear matter saturation point.
The most important parameter to be constrained is shown to 
be the symmetry energy slope $L$. Indeed, this parameter exhibits a linear correlation with the stellar radius, which is particularly clear for small mass
stars around  $1.0\;M_\odot$. The other equation of state parameters do not show clear correlations with the radius, within the present uncertainties.
Potential constraints on $L$, the  neutron star radius and  the equation of state from observations of thermal states of neutron stars are also discussed. 
The unified equations of state are made available in the supplementary material section and on the CompOSE database.
\end{abstract}
%%%%%%%%%%%%%%%%%

%%%%%%%%%%%%%%%%%
% insert suggested PACS numbers in braces on next line
% insert suggested keywords - APS authors don't need to do this
%%%%%%%%%%%%%%%%%
\maketitle

%%%%%%%%%%%%%%%%%%%%%%%%%
\section{Introduction}
\label{sect:introduction}
%%%%%%%%%%%%%%%%%%%%%%%%%
Simultaneous measurements of the masses and radii of neutron stars
(NS), if sufficiently precise, will impose strong constraints on the equation of state (EOS)  of dense matter significantly  above (standard)  nuclear  (baryon number) density $n_0=0.16~{\rm fm^{-3}}$ (corresponding to a mass-energy density $\rho_0=2.7\times 10^{14}~{\rm g~cm^{-3}}$). Both $n_0$ and $\rho_0$ are suitable units to measure   the baryon (number) density and mass-energy density in NS cores. In fact, the two most massive pulsars PSR J0348+0432 and PSR J1614$-$2230 alone,  with a mass close to $2\,M_\odot$ \cite{demorest,antoniadis}, already put quite stringent constraints on the EOS in the $5n_0 - 8n_0$ density range. These mass measurements are particularly relevant to assess the possible existence of exotic phases of dense matter  in the cores of massive NS. 

Big effort has been put into the determination of the radii of NS
but presently there is still a large uncertainty associated with this
quantity, see the discussion in \cite{P14,steinerEPJA2015,fortin2015}. Particularly interesting is the measurement of radii for the stellar mass range $1.3\;M_\odot  - 1.5\;M_\odot$, where on the one hand many precise  NS mass 
measurements  exist, and on the other hand dense matter theories predict a nearly constant value of $R$ (albeit different for various dense matter theories). We expect that up to  $2n_0 - 3n_0$ NS  matter involves nucleons only  and therefore that the radius for the ``canonical'' NS mass   $1.4\;M_\odot$, denoted usually as $R_{1.4}$  characterises the EOS  in the nucleon segment.  
Recently, a new constraint has been added to this discussion.  According to Ref. \cite{chen2015}, an EOS with  $M_{\rm max}>2\;M_\odot$ should produce $R_{1.4}\gtrsim 10.7$ km  in order to avoid being  non-causal at highest NS densities.  

We expect that future simultaneous determinations of
the mass and radius of a NS  with a 5\% precision will be possible
through the analysis of the X-ray emission of NS, thanks to the forthcoming
NICER \cite{nicer}, Athena \cite{athena+} and LOFT-like \cite{loft}
missions. It is therefore important to be able to quantify the
uncertainties introduced in the NS mass and radius calculations
by at the same time the  approximations used when constructing the complete EOS for stellar
matter and the scarce available constraints on the EOS at high densities,
large isospin asymmetries, or the lack of information about the
possible exotic states of the matter existing in the interior of a NS.

In the present work we aim at understanding how the calculation
of the NS radii are affected by the EOS
of the crust, having in mind that the  EOS  constructed
to describe NS matter are typically  {\it non-unified}, i.e. built
piecewise  starting from different models for each sector of NS
matter. This is to be contrasted with {\it unified} EOS, where all
segments  (outer crust, inner crust, liquid core) are calculated
starting from the same nuclear interaction. In practice, for  NS crust
with $\rho \lesssim 10^{11}~{\rm g~cm^{-3}}$ one uses experimental
nuclear masses. For higher crust densities, where the relevant
experimental nuclear masses are not available, they should be
calculated theoretically. Usually, one employs an effective nuclear
hamiltonian (or lagrangian) and a many-body method that makes the
calculation feasible  (typically the Thomas-Fermi approximation or 
the compressible liquid-drop model). It should be mentioned, that some
minor matching problems exist already at the transition between the experimentally based low-density segment of the EOS, and that obtained with an effective nuclear interaction, if the latter does not fit perfectly experimental nuclear masses. However,  examples in the present paper show that resulting uncertainty in $R$ is very small.  The calculated EOS  for the crust will depend on the assumed effective nuclear interaction, but the phase transition between the inner crust (including a possibility of a  bottom layer with nuclear pastas) and the liquid core will be described correctly. The EOS is then continuous through the whole NS core, and yields a unique  $R(M)$ for each effective interaction, with negligible residual model dependence.   

On the contrary, in the standard case of a non-unified EOS model, the resulting 
$R(M)$ depends on the procedure of  matching  the crust and core EOS segments.  As an example, in \cite{glendenning99} it is proposed that the Baym-Pethick-Sutherland (BPS) EOS  \cite{bps} is chosen to  describe the crust and a matching of the crust EOS to the core one is performed at $0.01\;{\rm fm}^{-3}$, while the core is described within a relativistic mean field (RMF) approach allowing for fitting several  parameters of nuclear
matter at  saturation.  Similarly a parametrization of the high-density equation of state
based on piecewise polytropes is presented in  \cite{Read} and allows to systematically
study the effect of  observational constraints on the EOS of cold stellar matter. Although for the high density range several models have been
considered, for low densities a single EOS, the one of Douchin and Haensel \cite{DHb} based on a specific Skyrme interaction, namely Sly4 \cite{DHa}, is used. 
In an
equivalent way, the authors of \cite{steinerEPJA2015} have studied constraints on the NS structure by considering two classes of EOS models, and in both the
BPS EOS was taken for the low density EOS, alone or supplemented by the
Negele-Vautherin EOS \cite{VN}. Both of these models are based on old energy functionals which do not fulfill present experimental nuclear physics constraints.
In all these examples, one can wonder by how much
 the simplified choice for the sub-nuclear density EOS affects the
conclusions obtained from experimental and observational
constraints on the EOS. In fact, in \cite{piekarewicz2014} it has been
argued that, depending on the assumed properties 
of the low density EOS, it is possible to obtain 
pressures at the crust-core transition large enough to explain  the large Vela
glitches, even considering the entrainment effect.
 This indicates that a proper description of the crust and the crust-core 
 transition, as well as a sensitivity study and a systematic  uncertainty evaluation are required. 
 
In the present paper,  we will first study how the matching of the crust EOS with the core one affects the NS radius and  the crust
thickness, when models that describe the crust and the core EOS are
not the same. In order to reduce the uncertainties introduced  on the
calculation of the star structure, some general indications will be
presented on how to build a non-unified EOS. 

Next we will take  a set of unified EOS obtained in the framework of 
the RMF models and Skyrme interactions. For both
frameworks we restrict ourselves to EOS that are able to describe a
$2\;M_\odot$ star and remain causal, a non-trivial condition for the
second set of non-relativistic models. 

In the case of the RMF models one can  consider also their extensions
allowing for the presence of hyperons. Vector-meson couplings to hyperons are obtained assuming the SU(6) symmetry. Repulsion in the hyperon sector associated with their coupling  to a  hidden-strangeness  vector-isoscalar meson $\phi$ allows for $M>2\;M_\odot$. We also study how adding the hidden-strangeness scalar-isoscalar meson $\sigma^\star$ to get a weak $\Lambda\Lambda$ attraction  softens the EOS. 
In principle the same exercise could be done for the non-relativistic models. However, the present uncertainties in the hyperon-nucleon and hyperon-hyperon interactions are such that the introduction of hyperon degrees of freedom is still extremely model-dependent. In particular, the most sophisticated many-body approaches available in the literature \cite{BHF} either did not yet succeeded in producing $2\;M_\odot$ stars, or cannot deal with the full baryonic octet \cite{pederiva}.
However even in the case of RMF, strong uncertainties are associated to the couplings. 
We make all the EOS 
used here available in the supplementary material section and on the  CompOSE database.\footnote{{\tt http://compose.obspm.fr}}  
 
Within our large set of unified EOS we will study  the dependence of the NS star radius and  the thickness of the crust on the mass in order to pin down the residual uncertainties due to our imperfect knowledge of the EOS parameters. As we remind in Section~II, the EOS of nuclear matter near $n_0$  and for small neutron excess is  constrained by the semi-empirical evaluations of  nuclear matter parameters extracted from nuclear physics data. We will seek for the correlations between theoretically calculated nuclear matter parameters near $n_0$ and NS structure. We will specifically show that the best correlation is obtained between the radius of light NS with $M\leq 1.4\;M_\odot$ and the symmetry energy slope $L$.
This confirms that indeed the $L$ parameter is the most important one to be constrained from laboratory experiments and/or ab-initio calculations. 
A most crucial constraint could potentially come from the threshold density above which the direct Urca (DUrca) process operates. Indeed the interval of $L$ which is compatible with terrestrial constraints largely overlaps with  
the one for which the nucleonic  DUrca process operates in massive NS. In turn, the presence of nucleonic DUrca appears to be needed in order to explain the thermal states of accreting neutron stars \cite{BY15}. This means that combining radii measurements with observations of thermal states of NS might constitute a very stringent test for the EOS.
 
 The plan of the paper is as follows. In Section~II we  give a very general overview of nuclear matter in NS. We also establish notations for nuclear matter  and its relation to the semi-empirical nuclear-matter parameters.  Section~III describes the different techniques that are used to match the crust and core EOS, and the resulting uncertainty associated to the star radius and the crust thickness. 
The relativistic and non relativistic unified EOS employed for this work are described in Section~IV, and the corresponding $M(R)$ relations are given. Section~V contains the main results of this work. The predictions for the radius and crust thickness are given, the correlation between the radius and the EOS parameters is discussed, and the different unified EOS are compared to the terrestrial constraints. Potential constraints from the necessity of DUrca processes to explain low-luminosity NS are presented. Finally Section~VI concludes the paper.  
%%%%%%%%%%%%%%%%%%%%%%%%%

\section{Nuclear matter in neutron stars  and semi-empirical parameters} 
%%%%%%%%%%%%%%%
Consider the NS interior from the very basic point of view of nuclear matter states relevant for each main NS layer. The $T=0$ approximation can be used since the Fermi energy of the nucleons is much larger than the thermal energy associated with the temperatures of $\sim 10^7-10^9$~K  expected inside NS. The outer core consists of a lattice of nuclear-matter droplets permeated by an electron gas. The inner crust is made of a lattice of nuclear-matter droplets coexisting with a neutron gas. With increasing pressure, droplets can become unstable with respect to  merging into infinite nuclear matter structures (rods, plates) immersed in a neutron gas. The plates of nuclear matter  then glue together leaving tubes filled with neutron gas, then the tubes break into bubbles of neutron gas in nuclear matter. Both the inner crust and the (possible) mantle of nuclear pastas form inhomogeneous two-phase states of nucleon matter. 

At the edge of the outer core, inhomogeneous nucleon matter coexists with uniform  homogeneous nuclear matter. To model it, we consider a mixture of strongly interacting neutrons and protons, with Coulomb interactions switched off. Let us define the baryon number density $n=n_n+n_p$ and the neutron excess parameter $\delta=(n_n-n_p)/n$. 
The energy per nucleon (excluding the nucleon rest energy) is $E_{_{\rm NM}}(n_{\rm
b},\delta)$.  Theoretical models of nuclear matter give $E_{_{\rm NM}}(n_{\rm
b},\delta)$ and yield a set of parameters that characterize the EOS  near the saturation point (minimum of $E_{_{\rm NM}}$)  and for small $\delta$.  For a given model, the
minimum of energy per nucleon, $E_{\rm s}$,  is reached at the saturation density $n=n_{\rm s}$ and for
$\delta=0$. 

The difference between the calculated  values for the saturation density $n_{\rm s}$ and the commonly used normal nuclear density $n_0$ defined in the first sentence of Section~I deserves a comment. The values of $n_{\rm s}$ are model-dependent and vary between  $0.146~{\rm fm^{-3}}$ and $0.154~{\rm fm^{-3}}$ for the RMF models (Table~\ref{tab:rmf}) and between $0.151~{\rm fm^{-3}}$ and $0.165~{\rm fm^{-3}}$ for the Skyrme models (Table~\ref{table:NNparam}). The use of a precise value of $n_{\rm s}$ is crucial for the correct calculation of the EOS. On the contrary, $n_0$ is just a chosen baryon number density unit. 

Let us define the so-called symmetry energy:
\begin{equation}
E_{\rm sym}\left(n\right)=\frac{1}{2}\left(\frac{\partial^2 E_{_{\rm NM}}}{\partial \delta^2}
\right)_{\delta=0}
\end{equation}
and its value at saturation: 
\begin{equation}
J=E_{\rm sym}\left(n_{\rm s}\right).
\end{equation}
Two additional parameters related to the first and second derivatives of the symmetry energy at the saturation point are respectively: the symmetry-energy slope parameter $L$, 
\begin{equation}
L=3 n_{\rm s}\left(\frac{\drom E_{\rm sym}}{\drom n}\right)_{n_{\rm s}}~,
\end{equation}
%%%%%%%%%
and the symmetry incompressibility  $K_{\rm sym}$:
%%%%%%%%
\begin{equation}
K_{\rm sym}=9n_{\rm s}^2\left(\frac{\drom^2 E_{\rm sym}}{\drom^2 n}
\right)_{n_{\rm s}}~.
\end{equation}
%%%%%%

Finally, the incompressibility at saturation $K$ is:
%%%%%%%%
\begin{equation}
K=
9n_{\rm s}^2\left(\frac{\partial^2 E_{_{\rm NM}}}{\partial n^2}
\right)_{n_{\rm s},\delta=0}~. 
\end{equation}
%%%%%%

The values of parameters $\lbrace n_{\rm s},E_{\rm s}, \ldots, K_{\rm sym}\rbrace$  for our sets of the  RMF and Skyrme models are given in Table~\ref{tab:rmf} and Table~\ref{table:NNparam}, respectively.

The knowledge of parameters $\lbrace n_{\rm s},E_{\rm s},K,J,\ldots,\rbrace$ is sufficient for reproducing theoretical  EOS of nuclear matter near the saturation point, a situation characteristic of laboratory nuclei. However, after being fine-tuned at the saturation point, the energy-density functionals are actually extrapolated up to $n\sim 8n_{\rm s}\simeq 8n_0$  and $\delta\simeq 1$, characteristic of the cores of massive NS. Therefore, making  $\lbrace n_{\rm s},E_{\rm s},K,J, \ldots,\rbrace$ consistent with the semi-empirical evaluations of these parameters obtained using a wealth of experimental data on atomic nuclei,  yields  constraints on the  corresponding EOS of NS, and consequently, NS models, and in particular - NS radii. 

%%%%%%%%%%%%%%%%%%%%%%%%%
\section{Non-unified equations of state and core-crust matching}

In the present section we will discuss the problem of the  core-crust
matching of the EOS when a non-unified EOS is used to describe stellar
matter. The use of a non-unified EOS will be shown to hardly
affect the determination of the NS mass but to have a significant influence on the radius
calculation.

\subsection{Different procedures for core-crust matching}
\label{subs:matching}

The determination of the mass and radius of a NS is possible
from the integration of the Tolman-Oppenheimer-Volkoff equations for spherical and
static relativistic stars \cite{tov}, given the EOS of stellar matter 
$P(\rho)$, where $P$ is the pressure and $\rho$ the mass-energy
density. The EOS for the whole NS is generally obtained by the matching of three
different segments:  the first one for the outer crust, the second one for the inner crust
and the last one for the core. The EOS for the outer crust, which extends from the
surface to the neutron drip density,  requires the knowledge of the masses of
neutron-rich nuclei \cite{bps,hp,ruester06}. This information comes
from experiments or, when
no information exists,  from some energy-density functional calculation. The
inner crust corresponds to a non-homogeneous region between the neutron drip and the
crust-core transition. This region may include non-spherical nuclear
clusters, generally known as pasta phases \cite{ravenhall83} and has
been described within several approaches \cite{horowitz04,horowitz05,maruyama05,sonoda,sonoda2,avancini08,newton09}. Finally the
core formed by a homogeneous liquid composed of neutrons, protons,
electrons, muons and possibly other exotic matter, in
$\beta$-equilibrium extends from the crust-core transition to the
center of the star. It should be pointed out, however, that in addition to exotic phases which can possibly appear at high densities, matter may also be non-homogenous in the core, e.g., in the form of a mixed hadron-quark phase \cite{YL14}.
In the present
 work we consider a homogeneous core.

Since the core accounts for most of the mass and radius of the star, 
authors frequently work with a non-unified EOS, and match
the core EOS to one for the crust, in particular the
Baym-Pethick-Sutherland (BPS) \cite{bps}
together with  the
Baym-Bethe-Pethick (BBP) \cite{bbp},  the
Negele-Vautherin (NV) \cite{NV} or the
Douchin-Haensel (DH) \cite{DHb}.  The matching is generally done so
that the pressure is an increasing function of the energy
density. This condition still leaves a quite large freedom in the
matching procedure.
In principle the matching procedures done at a specific density should be performed
using a Maxwell construction, i.e. at constant baryonic chemical
potential, so that the pressure is an increasing function of both the
density and the chemical potential. 

In the following a non-unified EOS is built from two different EOS. The one for the crust defined by $P_{\rm cr}$, $\rho_{\rm cr}$, $n_{\rm cr}$ is used up to $P_1$, $\rho_1$, $n_1$, while another one for the core: $P_{\rm co}$, $\rho_{\rm co}$, $n_{\rm co}$ is considered above $P_2$, $\rho_2$, $n_2$. The matching is performed in the region of pressure: $P_1\le P\le P_2$, and if $P_1\ne P_2$ a linear interpolation between $(P_1,\rho_1)$ and
   $(P_2,\rho_2)$ is considered. 
The pressures $P_1$ and $P_2$ are generally defined at a reference
density such as the neutron drip density $n_{\rm d}$, the crust-core
transition density $n_{\rm t}$, the saturation density $n_0$, and the 
density $n_{\rm c}$ where the two EOS cross.

\begin{figure}
\resizebox{\hsize}{!}{\includegraphics{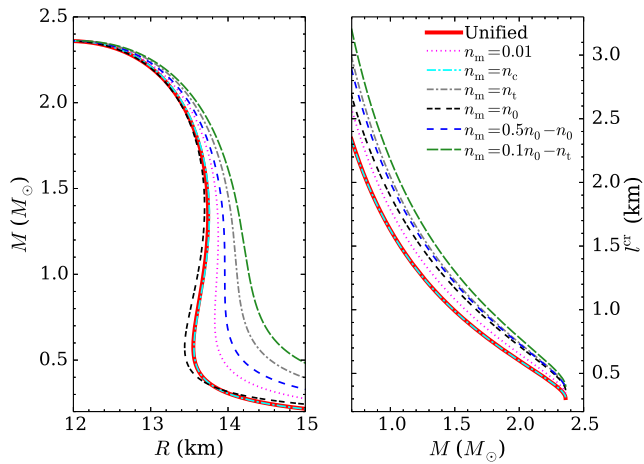}}
\caption{Mass versus radius (left) and crust thickness $l^{\rm cr}$ versus mass (right) for the relativistic mean field model GM1, using
  different matching procedures (see text).}
\label{fig:matching1}

\end{figure}

 In Fig.~\ref{fig:matching1} we plot the
radius-mass curves (left) and the crust thickness (right) versus the star mass obtained with the
GM1 parametrization with a purely nucleonic core obtained for different glueing procedures:
\begin{enumerate}
\item Unified:  by unified we
mean an EOS built with the DH EOS for the outer crust ($n\le 0.002$
fm$^{-3}$) and the inner crust
and core obtained within the same model, here GM1.
The inner crust was calculated within a
Thomas Fermi calculation of the pasta phase \cite{GP} and the
core EOS matches the inner crust at the crust-core transition
density $n_{\rm t}$;
\item $n_1=0.01$ fm$^{-3}$:  the crust BPS+BBP EOS
is glued to the core EOS at $0.01$ fm$^{-3}$ as indicated in
\cite{glendenning99};
\item $n_1=n_{\rm c}$: the glueing is done at the density
where the DH EOS and the core EOS cross as in \cite{Read};
\item $n_1=n_{\rm t}$: the  DH EOS is considered for the  crust and  homogeneous matter EOS for
  $n>n_{\rm t}$.
\item $n_1=n_0$: the DH  EOS is used for $n<n_0$ and the core EOS
  above the saturation density $n_0$.
\item $n_1=0.5n_0$, $n_2=n_0$: DH  EOS is used for $n<0.5n_0$, the homogeneous matter
  EOS is used above $n_0$.
\item $n_1=0.1n_0$, $n_2=n_{\rm t}$: a low matching of the EOS is considered. The
  DH EOS is used for
  $n<0.1n_0$ and the core EOS above $n_{\rm t}$.
\end{enumerate}
 If the matching is defined
at a given density $n_{\rm m}=n_1$, the glueing is done
imposing $P_2=P_1$.
The curves do not coincide because
the matching has been performed in different ways. While the
 maximum mass allowed for a stable star is not affected by the
 crust-core matching chosen, the same is not true for the radius and
 crust thickness of stars with a standard mass of $\sim 1.4\,
 M_\odot$.  
 The two EOS considered in this
example for the crust and the core have quite different
properties at saturation density, in particular for the
density dependence of the symmetry energy, see Table
\ref{tab:rmf}. This situation is, however,  common in the literature. In fact, the GM1
 EOS \cite{GM1} was parametrized to describe both nuclear saturation
 properties and neutron star properties. 

 In Table~\ref{tab:delta}, the  radius and crust thickness
 of $1.0$ and $1.4\, M_\odot$ NS are given for three models,
 GM1, NL3 \cite{NL3} and NL3$\omega\rho$ \cite{NL3wra}, and several matching
 schemes, together with relative differences with respect to the value
 for the unified EOS.
 
 As expected the crust-core matching is affecting more
 strongly the less massive stars. Depending on the matching procedure
 the difference in the radius and the crust thickness for a $1.0\;
 M_\odot$ star can be as large as $\sim 1$~km and  $\sim 0.5$~km, respectively. 
This corresponds to relative differences as large as $\sim 4\%$ for the radius and $\sim 30\%$ for the crust thickness. This is to be compared with the expected precision of $\sim 5\%$ on the radius measurement from future X-ray telescopes (NICER, Athena, \ldots).
Similarly the crust thickness differs by $\sim 0.5$~km depending on the glueing. This quantity is particularly important for the study of the thermal relaxation of accreting NS \cite{SY07,PR13}, the glitch phenomenon \cite{AG12,Pi14}, the torsional crustal vibrations and the maximum quadrupole ellipticity sustainable by the crust \cite{GN11}. 

The differences between matchings are much smaller if the NL3$\omega\rho$ core
 EOS is considered, because this model has nuclear matter saturation properties
 similar to the ones of the Sly4 parametrization \cite{DHa} used in the DH EOS.

\begin{table}
\begin{ruledtabular}
\center\begin{tabular}{lcccccccccccc}
 &$R_1$ &$\Delta R_1$ &$R_{1.4}$&$\Delta R_{1.4}$&$l^{\rm cr}_1$&$\Delta l^{\rm cr}_1$& $l^{\rm cr}_{1.4}$ & $\Delta l^{\rm cr}_{1.4}$ \\
\hline
\hline
GM1\\
\hline
unified           &13.71 &-     &13.76 &-     &1.62 &-     &1.09 &-    \\
\hline
$n=0.01$       &13.86 &1.09  &13.86 &0.73  &1.78 &9.88  &1.19 &9.17 \\
$n_{\rm t}$          &14.12 &2.99  &13.92 &1.16  &1.64 &1.23  &1.10 &0.92 \\
$n_0$          &13.61 &-0.73 &13.70 &-0.44 &2.04 &25.93 &1.36 &24.77\\
0.5$n_0-n_0$&13.96 &1.82  &13.92 &1.16  &2.00 &23.46 &1.33 &22.02\\
0.1$n_0-n_{\rm t}$&14.27 &4.08  &14.12 &2.62  &2.18 &34.57 &1.44 &32.11\\
Max. diff. &0.66  &-     &0.42  &-     &0.56 &-     &0.35 &-    \\
\hline
\hline

NL3\\
\hline
unified& 14.54 &-& 14.63&-& 1.91&-&1.30&-\\
\hline
$n=0.01$&14.78&1.65&14.78&1.03&2.15&12.57&1.45&11.54\\
$n_{\rm c}$&14.97&2.96&14.91&1.91&2.35&23.04&1.58&21.54\\
$n_{\rm t}$& 14.96&2.89&14.90&1.85&2.34&22.51&1.57&20.77\\
$n_0$&14.00&-3.71&14.26&-2.53&2.02&5.76&1.42&9.23\\
0.5$n_0-n_0$&14.47&-0.48&14.57&-0.41&2.17&13.61&1.50&15.38\\
0.1$n_0-n_{\rm t}$&15.09&3.78&14.97&2.32&2.46&28.80&1.65&26.92\\
Max. diff. & 1.09&-&0.71&-&0.55&-&0.35&-\\
\hline
\hline

NL3$\omega\rho$\\
\hline
unified&13.42&-&13.75&-&2.02&-&1.43&-\\
\hline
$n=0.01$&13.51&0.67&13.81&0.44&2.11&4.46&1.49&4.20\\
$n_{\rm c}$&13.5&1.12&13.85&0.73&2.18&7.92&1.53&6.99\\
$n_{\rm t}$& 13.5&0.60&13.8&0.36&2.1&3.96&1.48&3.50\\
$n_0$&13.49&0.52&13.8&0.36&2.1&3.96&1.48&3.50\\
0.5$n_0-n_0$&13.51&0.67&13.81&0.44&2.11&4.46&1.49&4.20\\
0.1$n_0-n_{\rm t}$&13.49&0.52&13.8&0.36&2.1&3.96&1.48&3.50\\
Max. diff.&0.15&-&0.10 &-&0.16 &-&0.10&-\\

\end{tabular}
\end{ruledtabular}

\caption{NS radii $R_1$ and $R_{1.4}$ (in km)  and crust thicknesses $l^{\rm cr}_1$ and $l^{\rm cr}_{1.4}$ (in km) for masses of $1.0$ and $1.4\;M_\odot$ for different matchings between the core and the crust. $\Delta x$ (in $\%$) for a given quantity $x$ corresponds to the relative difference between the value of $x$ for unified EOS and the one for a given matching. Three functionals are considered: NL3, 
  NL3$\omega\rho$ and GM1.}
\label{tab:delta}
\end{table}
\subsection{Thermodynamic consistency}
\label{sect:tc}
Two basic methods can be used in order to match two EOS for the crust and the core: the first based on the $P(n)$ relation and the second on the $P(\rho)$ function.\\

The first method consists in treating the baryon number density as an independent variable. 
Consider an EOS for the crust, $P_{\rm cr}(n)$ and $\rho_{\rm cr}(n)$, and another one for the core, $P_{\rm co}(n)$ and $\rho_{\rm co}(n)$. 

Let us assume that the matching region lies between two densities, $n_1$ and $n_2>n_1$. First let us build the matched $P(n)$ function. For $n< n_1$, $P(n)=P_{\rm cr}(n)$ and for $n> n_2$, $P(n)=P_{\rm co}(n)$. In the matching region, one can assume a form (usually linear or logarithmic) for the function $P(n)$ such that
$P(n_1)=P_1=P_{\rm cr}(n_1)$ and $P(n_2)=P_2=P_{\rm co}(n_2)$. 

Then one needs to build the function $\rho (n)$. For $n<n_1$, $\rho(n)=\rho_{\rm cr}(n)$. 
Let us define the chemical potential at the density $n_1$: $\mu_1=(P_1+\rho_{\rm cr}(n_1))/n_1$.  By imposing thermodynamic consistency the 
value of chemical potential $\mu$  at a density $n$ in the matching region can be derived using the $P(n)$ relation:
\begin{equation}
 \mu(n)=\mu_1+\int_{n_1}^n\frac{\drom P(n)}{n}.
\label{eqn:mudef}
\end{equation}
Finally the matched mass-energy density is 
\begin{equation}
 \rho (n)= n\mu(n)-P(n).
\label{eqn:rhodef}
\end{equation}
However this technique generally leads to thermodynamic inconsistency: the value of chemical potential $\mu_2$ at the density $n_2$ obtained from Eq.~(\ref{eqn:mudef}) differs from the one given by the core EOS: $\mu_{\rm co}(n_2)=(P_2+\rho_{\rm co}(n_2))/n_2$. As a consequence $\rho (n_2)$ given by Eq.~(\ref{eqn:rhodef}) is different from $\rho_{\rm co}(n_2)$.
In order to get a thermodynamically consistent EOS for $n>n_2$ one has to add a constant value (a mass-energy shift):
\begin{equation}
 \Delta\mu= \mu(n_2)-\mu_{\rm co}(n_2)
\end{equation}
%%%%%%%%%%%%%%%%
to the chemical potential in the core. Then the mass-energy density $\rho(n)$ for $n>n_2$ is 
\begin{equation}
 \rho(n)=\rho_{\rm co}(n)+n \Delta\mu.
\end{equation}
%%%%%%%%%%%%%%%%
Of course, such a procedure affects the whole EOS for the core - but the main effect on the  $M(R)$ relation is for NS with a central pressure close to $P_2$.\\

The second method considers the mass-energy density $\rho$ as an independent variable. This can be motivated by the TOV equations since this quantity and the function $P(\rho)$ actually enter the stress-energy tensor  in the  Einstein equations. 
Thus the EOS can be written in the form: $P_{\rm cr}(\rho)$ and $n_{\rm cr}(\rho)$ for the crust, and: $P_{\rm co}(\rho)$ and $n_{\rm co}(\rho)$ for the core. The matching region is defined such that $\rho_1< \rho < \rho_2$. 

The first step consists in obtaining the function $P(\rho)$. For $\rho< \rho_1$, $P(\rho)=P_{\rm cr}(\rho)$ and for $\rho> \rho_2$, $P(\rho)=P_{\rm co}(\rho)$. Similarly to the first method one can assume a form for the function $P(\rho)$ in the matching region such that $P(\rho_1)=P_1=P_{\rm cr}(\rho_1)$, $P(\rho_2)=P_2=P_{\rm co}(\rho_2)$.

Then one wants to derive the relation $n(\rho)$. For $\rho< \rho_1$ one has $n(\rho)=n_{\rm cr}(\rho)$. Let us define $n_1=n_{\rm cr}(\rho_1)$. Assuming thermodynamic consistency, in the matching region, i.e. $\rho_1\leq \rho < \rho_2$, one gets 
\begin{equation}
 n(\rho)=n_1 \exp\left({\int_{\rho_1}^\rho\frac{\drom \rho}{P(\rho)+\rho}}\right).
\end{equation}
However, as for the first method, this construction does not ensure that $n(\rho_2)$ obtained from the previous formula is equal to $n_{\rm co}(\rho_2)$. A similar conclusion can be reached for the chemical potential at $\rho_2$. Thus one has to modify the $n(\rho)$ dependence for 
the core EOS in order to ensure thermodynamic consistency. For $\rho>\rho_2$, the matched EOS is 
\begin{equation}
n({\rho})=n_{\rm co}(\rho) \frac{n(\rho_2)}{n_{\rm co}(\rho_2)}.
\label{eqn:nmatched}
\end{equation}
This approach does not affect the $P(\rho)$
relation (nor the gravitational mass and the radius), but strictly speaking the microscopic model of dense matter is changed
since it is the baryon number density which is the basic quantity for the theoretical calculations, within the many-body theory, of dense matter properties. Of course the accepted procedure given by Eq.~\ref{eqn:nmatched} also influences the value of a baryon chemical potential (dividing it by the same factor).\\

\subsection{Allowed region for the crust-core matching}
\begin{figure}[h]
\resizebox{\hsize}{!}{\includegraphics{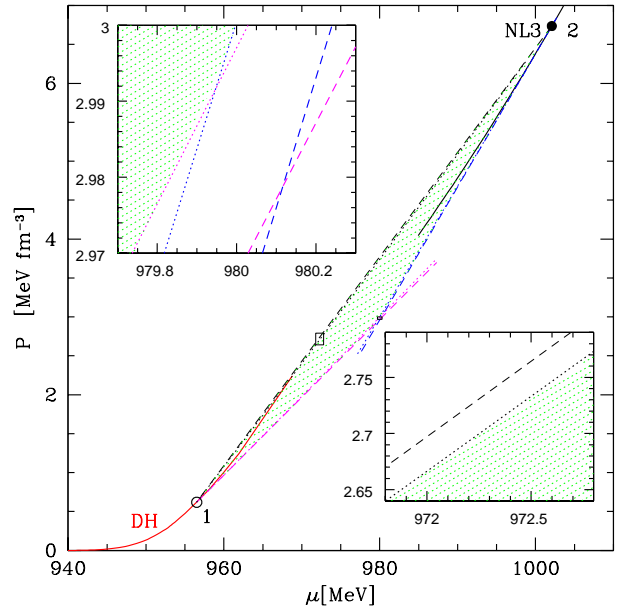}}
\caption{Pressure $P$ versus chemical potential $\mu$, for the NL3 EOS for the core (black solid line) 
and DH for the crust EOS (red solid line). The presented situation corresponds to a matching between
$n_1=0.09\,{\rm fm}^{-3}$ and $n_2=n_0=0.16$ fm$^{-3}$.
The dashed lines correspond to the condition of thermodynamical consistency and are given by Eqs.~(\ref{eq:inctan}-\ref{eq:inccon}). 
The dotted lines are given by the causality limit: Eqs.~(\ref{eq:causal}-\ref{eq:caucon}).
The area defined by these lines corresponds to the shaded region and is a bit smaller than the region 
allowed for a thermodynamically consistent matching between points {\bf 1} and {\bf 2} (see inserts).}
\label{fig:pmu}
\end{figure}
In principle when glueing two EOS, one should match all thermodynamic quantities: the
pressure $P$, the energy density $\rho$, and the baryonic density $n$. In other words,
a pair of functions for the pressure and the energy density  should be
constructed so that the thermodynamic
consistency is fulfilled. 

Let us consider the EOS for the core and the crust, this time in terms of the chemical potential $\mu$, $P_{\rm cr}(\mu)$ in the crust and $P_{\rm co}(\mu)$ in the core. The matching region is defined by $\mu_1<\mu<\mu_2$. Let us define $P_{\rm cr}(\mu_1)=P_1$ and $P_{\rm co}(\mu_2)=P_2$.
The function $P(\mu)$ in the matching region and its first derivative, which is the baryon
number density $n$, should  fulfill the conditions of continuity given by 
%%%%%%%%%%%%%%%%%%%%%
\begin{equation}
P(\mu_1)=P_1 , ~~ 
P(\mu_2)=P_2~.
\end{equation}
%%%%%%%%%%%%%%%%%%%%%%%%%%%%

\begin{figure}[h]
\resizebox{\hsize}{!}{\includegraphics{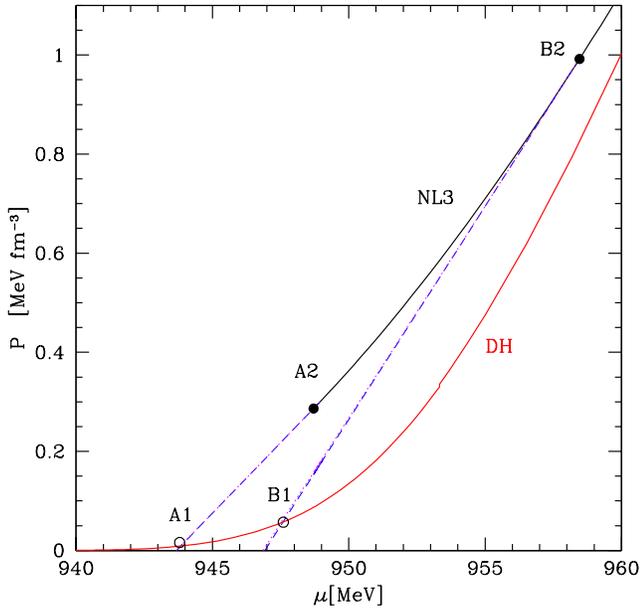}}
\caption{Pressure $P$ versus chemical potential $\mu$, for the NL3 EOS for the core (black solid line) 
and DH for the crust EOS (red solid line). The points A2 and B2 correspond to two different values of $n_2$: $n_{\rm t}$, the core-crust transition density and $n_0/2$, respectively. 
The dashed lines indicate
the condition of thermodynamical consistency and the dotted lines mark the causality limit. 
They are almost indistinguishable.
The points A1 and B1 correspond to the higher limits on 
$\mu$ or equivalently $n$, such that a thermodynamically consistent glueing with the core at the points A2 and B2 exists.}
\label{fig:pmub}
\end{figure}
Thermodynamic consistency and causality imply that  the following  conditions on the  derivatives must  be fulfilled in the matching region:
\begin{enumerate}
 \item $n$ is an increasing function of $P$, i.e. $P(\mu)$ is increasing and convex;
 \item $\left({\drom P}/{\drom\rho}\right)^{1/2}=v_{\rm sound}/c\le 1$, with the mass-energy density $\rho(\mu)=n(\mu)\mu-P(\mu)$.
\end{enumerate}
From the first requirement one can derive a necessary condition (using Lagrange's mean value theorem):
\begin{equation}
 n_1<\frac{P_2-P_1}{\mu_2-\mu_1}<n_2,\\
\label{eqn:Lagrange}
\end{equation}
\begin{equation}
{\rm with}~~n_1=\left(\frac{{\rm d}P_{\rm cr}}{{\rm d}\mu}\right)_{\mu_1} ~~ {\rm and} ~~n_2=\left(\frac{{\rm d}P_{\rm co}}{{\rm d}\mu}\right)_{\mu_2}\nonumber.
\end{equation}

If the above inequality is not fulfilled, a matching of the crust and core EOS using a continuous $P(\mu)$ function cannot be obtained.

Fig.~\ref{fig:pmu} shows an example of the matching between the DH EOS for the crust and the NL3 EOS for the core, with $n_1=0.09\,{\rm fm}^{-3}$ 
and $n_2=n_0$. The points {\bf 1} and {\bf 2} have the coordinates $\left(n_1,P_1(\mu_1)\right)$ and $\left(n_2,P_2(\mu_2)\right)$, respectively.

Any thermodynamically consistent EOS is 
located in the triangle defined by the two tangents at the points 1 and 2, 
\begin{equation}
 P=P_1+n_1 (\mu-\mu_1),~~~~~P=P_2+n_2 (\mu-\mu_2)
 \label{eq:inctan}
\end{equation}
and the straight line connecting these two points, 
\begin{equation}
 P=P_1+(P_2-P_1)\frac{\mu-\mu_1}{\mu_2-\mu_1}.
 \label{eq:inccon}
\end{equation}
However the bounds defined by Eqs.~(\ref{eq:inctan}-\ref{eq:inccon}) describe incompressible matter
with a constant baryon number density equal to $n_1$, $n_2$, $(P_2-P_1)/(\mu_2-\mu_1)$, respectively.

The additional constraint resulting from the causality requirement reduces the allowed region, but the change
is very small (see zoomed inserts in Fig.~\ref{fig:pmu}). Instead of the tangents at the points
{\bf 1} and {\bf 2} given by Eq.~(\ref{eq:inctan}) the causality limit corresponds to 
\begin{equation}
 P=P_i+n_i (\mu-\mu_i) \frac{\mu+\mu_i}{2\mu_i}~~~~i=1,2.
 \label{eq:causal}
\end{equation}
The line connecting the two points and fulfilling the causality condition is 
\begin{equation}
 P=P_1+(P_2-P_1) \frac{\mu^2-\mu_1^2}{\mu_2^2-\mu_1^2}.
 \label{eq:caucon}
\end{equation}

In Fig.~\ref{fig:pmub} the matching conditions for the same EOS as in Fig.~\ref{fig:pmu} but for lower $n_2$ (or equivalently $\mu_2$) are presented.
The points A2 and B2 correspond to $n_2=n_{\rm t}$, the transition density between the core and the crust, and $n_2=n_0/2$,
respectively.
In these cases, for a given $n_2$ two upper limits on $n_1$ can be obtained: the first one by the crossing point between the tangent at the point 2 given by Eq.~(\ref{eqn:Lagrange}) and the crust EOS, and the second one by the intersection of the line defined by Eq.~(\ref{eq:causal}) and the crust EOS. Here both upper limits are actually almost identical and correspond to the points A1 and B1, obtained for the points A2 and B2, respectively.  

For the matching of the NL3 EOS with the DH crust 
if we choose $n_2=n_0/2$ (point $B_2$ in the figure) then the condition given 
by Eqs.~(\ref{eqn:Lagrange}-\ref{eq:causal}) 
results in $n_1<0.02$~fm$^{-3}$ (point $B_1$). 
It means that the matching region in term of $n$ and $P$ should be extremely large, with $n_2>4n_1$ and $P_2>16P_1$.

A similar estimation for the point $A_2$ with $n_2=n_{\rm t}$ results in $n_1<0.0075$~fm$^{-3}$ (upper limit marked by the point $A_1$) and the matching described
in Section~\ref{subs:matching} for NL3 model ($n_1=0.1n_0$, $n_2=n_{\rm t}$) cannot be performed in a thermodynamically consistent and causal way, unless one changes the core EOS,
as described in Section~\ref{sect:tc}. As a consequence not all matchings presented in Section~\ref{sect:tc} are thermodynamically consistent and/or causal.
%%%%%%%%%%%%%%%%%%%%
\section{Unified equations of state}
\label{sec:unified}
We introduce a set of unified EOS which were
built within a RMF approach or using  non-relativistic Skyrme
interactions. The choice of models takes into account the astrophysical constraints on
the maximum NS mass and the speed of sound,
%%%%%%%%%%%%%%%%%%%%%
\begin{itemize}
\item $M_{\rm max}\ge 2\,M_{\odot}$;
\item $v_{\rm sound}(2\,M_{\odot})< c$.
\end{itemize}
%%%%%%%%%%%%%%%%%%%
To these two constraints we will add 
 experimental and theoretical constraints on nuclear matter properties
 and we will
discuss the uncertainty on the determination of the radius and the
crust thickness of $1.0\, M_\odot$, $1.4\, M_\odot$ and $1.8\, M_\odot$ stars.

Some of the proposed EOS, namely the RMF EOS  are not 
fully unified since the outer crust is not calculated within the
framework of the model that defines the rest of the EOS, but we have
checked that since most of the outer crust is defined by experimental
results, the use of other EOS for the outer crust, such as \cite{hp,ruester06}, does not
significantly affect the star radius with a mass above $1.0\;M_\odot$.

\subsection{RMF unified EOS}

\begin{table*}[tb]
\begin{ruledtabular}
\center\begin{tabular}{lccccccclcccl}
Model& $n_{\rm s}$       & $E_{\rm s}$ & $K$  & $J$  & $L$ &$K_{\rm sym}$&$n_{\rm t}$&Pasta&$M_{\rm max}^{\rm noY}$ & $n_{\rm DU}$
&$M_{\rm DU}$& Ref\\
       & (fm$^{-3}$) &(MeV) & (MeV)& (MeV)&(MeV)& (MeV)       &(fm$^{-3}$)  &     &  $(\msun)$&(fm$^{-3}$)&  $(\msun)$&\\
\hline
NL3                 & 0.149 & -16.2 & 271.6 & 37.4 & 118.9 & 101.6 &0.056& d       &   2.77 & 0.20 & 0.84& \cite{NL3}\\
NL3${\omega \rho}$ & 0.148 & -16.2 & 271.6 & 31.7 &  55.5 & -7.6  &0.082& s, r, d &   2.75 & 0.50 & 2.55 & \cite{NL3wra}\\
DDME2               & 0.152 & -16.1 & 250.9 & 32.3 &  51.2 & -87.1 &0.072& s, r, d &   2.48 & 0.54 & 2.29 & \cite{DDME2}\\
GM1                 & 0.154 & -16.3 & 300.7 & 32.5 &  94.4 &  18.1 &0.064& d       &   2.36 & 0.28 & 1.10& \cite{GM1}\\ 
TM1                 & 0.146 & -16.3 & 281.2 & 36.9 & 111.2 &  33.8 &0.058& d       &   2.18 & 0.21 & 0.81& \cite{TM1}\\
DDH$\delta$         & 0.153 & -16.3 & 240.3 & 25.6 &  48.6 &  91.4 &0.080& s, r, d &   2.14 & 0.44 & 1.54& \cite{G04}\\
DD2                 & 0.149 & -16.0 & 242.6 & 31.7 &  55.0 & -93.2 &0.067& s, r, d &   2.42 & 0.54 & 2.18 & \cite{BH14}\\
BSR2                & 0.149 & -16.0 & 239.9 & 31.5 &  62.0 &  -3.1 &0.065& s,r,d        &   2.38 & 0.37 & 1.61 & \cite{DK07,A10}\\
BSR6                & 0.149 & -16.1 & 235.8 & 35.6 &  85.7 & -49.6 &  0.061   &  d       &  2.44  & 0.27 & 1.00    & \cite{DK07,A10} \\
\end{tabular}
\caption{Nuclear and astrophysical properties of the RMF models. Saturation density $n_{\rm s}$, energy per nucleon at saturation  $E_{\rm s}$, incompressibility
for symmetric nuclear matter (at saturation) $K$, symmetry energy (at saturation) $J$, symmetry energy slope parameter $L$, and symmetry incompressibility $K_{\rm sym}$. All parameters calculated at 
the nuclear matter saturation point. The density at the edge of the liquid 
uniform core is denoted as $n_{\rm t}$. In the column "Pasta", the type(s) of pasta phase in the bottom layer (mantle) above the edge of the core is indicated: s stands for slab, r for rod,  and d
for droplet phases. $M_{\rm max}^{\rm noY}$ is the maximum mass for a purely nucleonic core composition. $n_{\rm DU}$ and $M_{\rm DU}$ are respectively the baryon density and NS mass threshold above which the nucleonic DUrca is switched on for a purely nucleonic core.}
\label{tab:rmf}
\end{ruledtabular}
\end{table*}

In the present study we consider  two different types of models within
the relativistic mean field (RMF) approach: (i) non-linear Walecka models (NLWM) with constant coupling parameters, and (ii) density-dependent hadronic models (DDH) with density-dependent coupling parameters. 
The only condition that has been
imposed {\it a priori} is that the models describe a   $2\;M_\odot$ star.
Within the first category a set of models that span a quite large
range of nuclear saturation properties was chosen: NL3 \cite{NL3} with a large symmetry energy slope and incompressibility at saturation and which was fitted to the ground state properties of both stable and unstable nuclei,
NL3$\omega\rho$ \cite{NL3wra} which, compared to NL3, has a softer
density dependence of the symmetry energy through the inclusion of a
non-linear $\omega\rho$ term, GM1 \cite{GM1} fitted to describe
nuclear matter saturation properties subject to NS 
mass-radius constraints, TM1
\cite{TM1}  which includes
non-linear $\omega$
meson terms in order to soften the EOS at high densities and is the EOS of
one of the classical supernova EOS \cite{stosa,stosb}, and two
paramerizations, BSR2 and BSR6, with several non-linear terms mixing
the $\omega, \, \rho$ and $\sigma$ mesons \cite{DK07,A10}. Within the second type, three EOS were considered:
DDME2 \cite{DDME2}, DD2 \cite{typel2009} and DDH$\delta$ \cite{G04},
the last one also including the
$\delta$ meson. Some properties of the set of models we use are
indicated in Table~\ref{tab:rmf}.

We have built unified EOS for these models in the following
way: a) for the outer crust we  take the  EOS proposed in \cite {ruester06}; b) for
the inner crust we perform a Thomas Fermi calculation and allow for
non-spherical clusters according to \cite{GP0,GP}; c) for the core we
consider the homogeneous matter EOS. The transition between the inner
crust and the core is smooth. The maximum mass stars in Table
\ref{tab:rmf} have been obtained with the unified EOS.

Two compositions are considered: purely nucleonic and
baryonic matter with both nucleons and hyperons. 
The nucleonic models, the so-called {\bf noY} models, include the
scalar $\sigma$, vector $\omega$, and vector-isovector $\rho$ meson
fields (possibly also the $\delta$ meson) together with the nucleon doublet: neutron $n$ and proton $p$. 
The  {\bf Y} and {\bf Yss} models denote hyperonic EOS and, with
respect to the {\bf noY} models, they also include the six
lightest hyperons ($\Lambda^0$, the $\Sigma^+,\Sigma^0,\Sigma^-$ triplet, and the 
$\Xi^0, \Xi^-$ doublet) and  the hidden-strangeness  vector-isoscalar
$\phi$ meson  for the {\bf Y} models, or  the $\phi$ meson together with the hidden-strangeness
scalar-isoscalar $\sigma^*$ for the {\bf Yss} models.

The vector meson-hyperon  coupling constants  are always calculated assuming SU(6) symmetry (see eg. \cite{MC13,SA12,BH12}):
\begin{eqnarray}
	&&\frac{1}{3} g_{\omega N} = \frac{1}{2} g_{\omega\Lambda} = \frac{1}{2} g_{\omega\Sigma} = g_{\omega\Xi} \nonumber \ ,\\
	&&2 g_{\phi\Lambda} = 2 g_{\phi\Sigma} = g_{\phi\Xi} =- \frac{2\sqrt{2}}{3} g_{\omega N} \nonumber\ ,\\
	&&g_{\rho N} = \frac{1}{2} g_{\rho\Sigma} = g_{\rho\Xi} \ ,\\
    &&g_{\phi N} = 0 \nonumber\ ,\\
	&&g_{\rho\Lambda} = 0 \nonumber\ ,
	\label{eq:SU6-relation}
\end{eqnarray}
where $N$ stands for nucleons.
The $g_{\sigma Y}$ couplings, where $Y$ stands for hyperons ($Y=\Lambda,\Sigma,\Xi$), are obtained from the hyperon potential in symmetric nuclear matter, $U_{Y}^{(N)}$,
\begin{equation}
	U_{Y}^{(N)} =
	- g_{\sigma Y} s_0 
	+ g_{\omega Y} w_0 \,
	\label{eq:potential}
\end{equation}
with $s_0$ and $w_0$  the mean-field values of the $\sigma$ and $\omega$ meson fields, respectively.
Here we adopt the following values at saturation density: $U_{\Lambda}^{(N)}(n_{\rm s}) = -28$ MeV,
$U_{\Xi}^{(N)}(n_{\rm s}) = -18$ MeV, and $U_{\Sigma}^{(N)}(n_{\rm s}) = 30$ MeV \cite{SBG00} (see also discussion in \cite{OP15}).

In the {\bf Yss} model the $\sigma^{\ast}$ meson is also included. It is assumed that it does not couple to a nucleon, i.e. $g_{\sigma^{\ast}N}=0$. The $\Lambda$-potential in $\Lambda$-matter is given by:
\begin{equation}
	U_{\Lambda}^{(\Lambda)} =
	- g_{\sigma \Lambda} s_0
	- g_{\sigma^{\ast} \Lambda} s_0^{\ast}
	+ g_{\omega \Lambda} w_0
	+ g_{\phi \Lambda} f_0 \ . 
	\label{eq:potential-ss}
\end{equation}
with $s_0^{\ast}$ and $f_0$ the mean-field values of the $\sigma^{\ast}$ and $\phi$ meson fields, respectively.
Taking  $U_{\Lambda}^{(\Lambda)}(n_{\rm s}) =-5$ MeV \cite{T01} (see also discussion in \cite{OP15}), the value of $g_{\sigma^{\ast} \Lambda}$ can then be fixed. The two remaining coupling constants can be derived taking $g_{\sigma^{\ast}\Sigma}=g_{\sigma^{\ast}\Lambda}$ and assuming that $U_{\Xi}^{(\Xi)} \simeq 2 U_{\Lambda}^{(\Lambda)}$. The $\Xi$-potential, $U_{\Xi}^{(\Xi)}$, in symmetric $\Xi^0-\Xi^-$ matter is given by an expression similar to Eq.~(\ref{eq:potential-ss}) replacing $\Lambda$ by $\Xi$.
For DDH$\delta$ we only present the  {\bf Y} results because even
in this case the maximum mass is far from $2\;M_\odot$, and the
presence of the
$\delta$-meson brings extra unknowns in the definition of the
hyperon-meson couplings.

\begin{figure}
%\begin{tabular}{cc}
\includegraphics[width=0.78\linewidth]{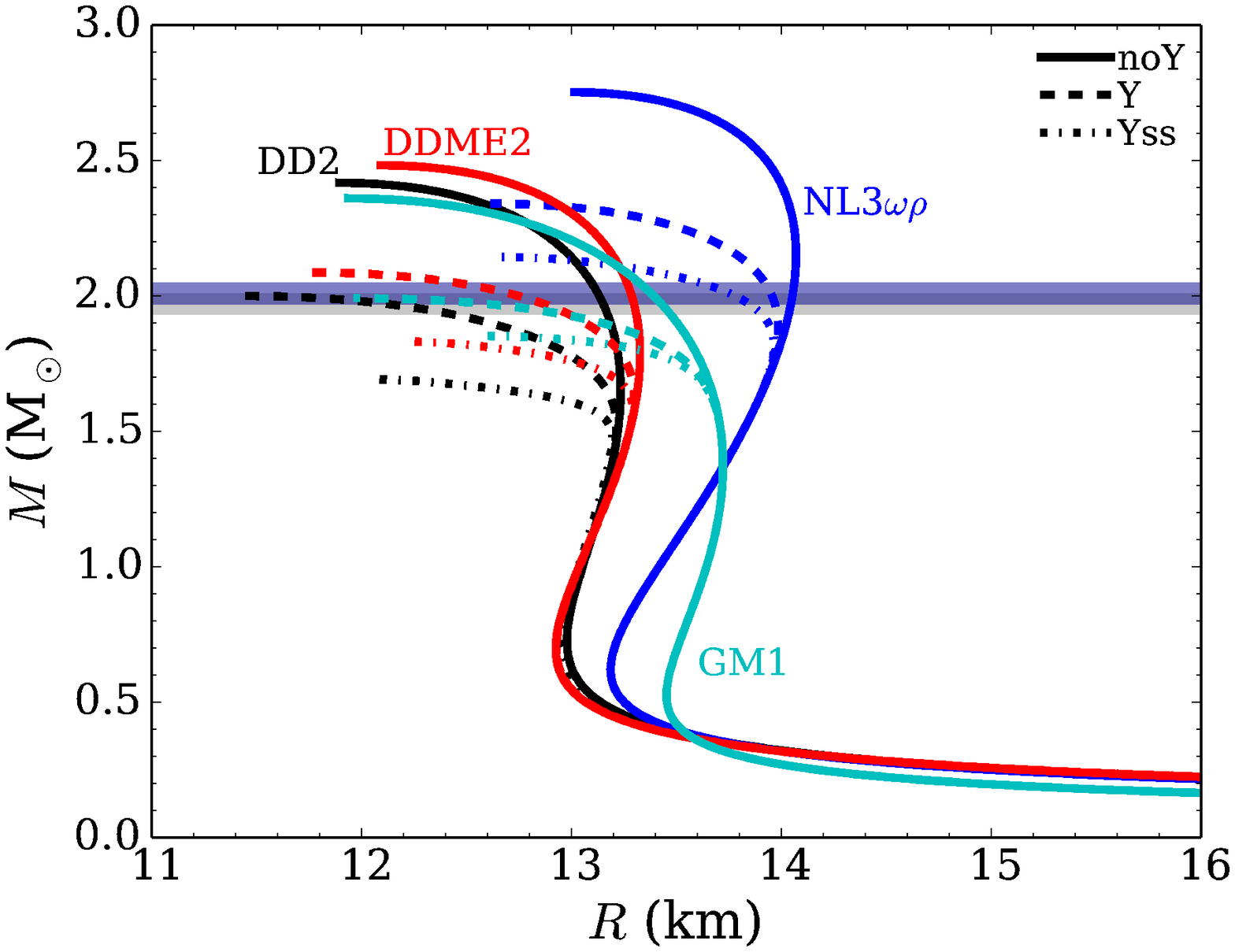}\\
\includegraphics[width=0.78\linewidth]{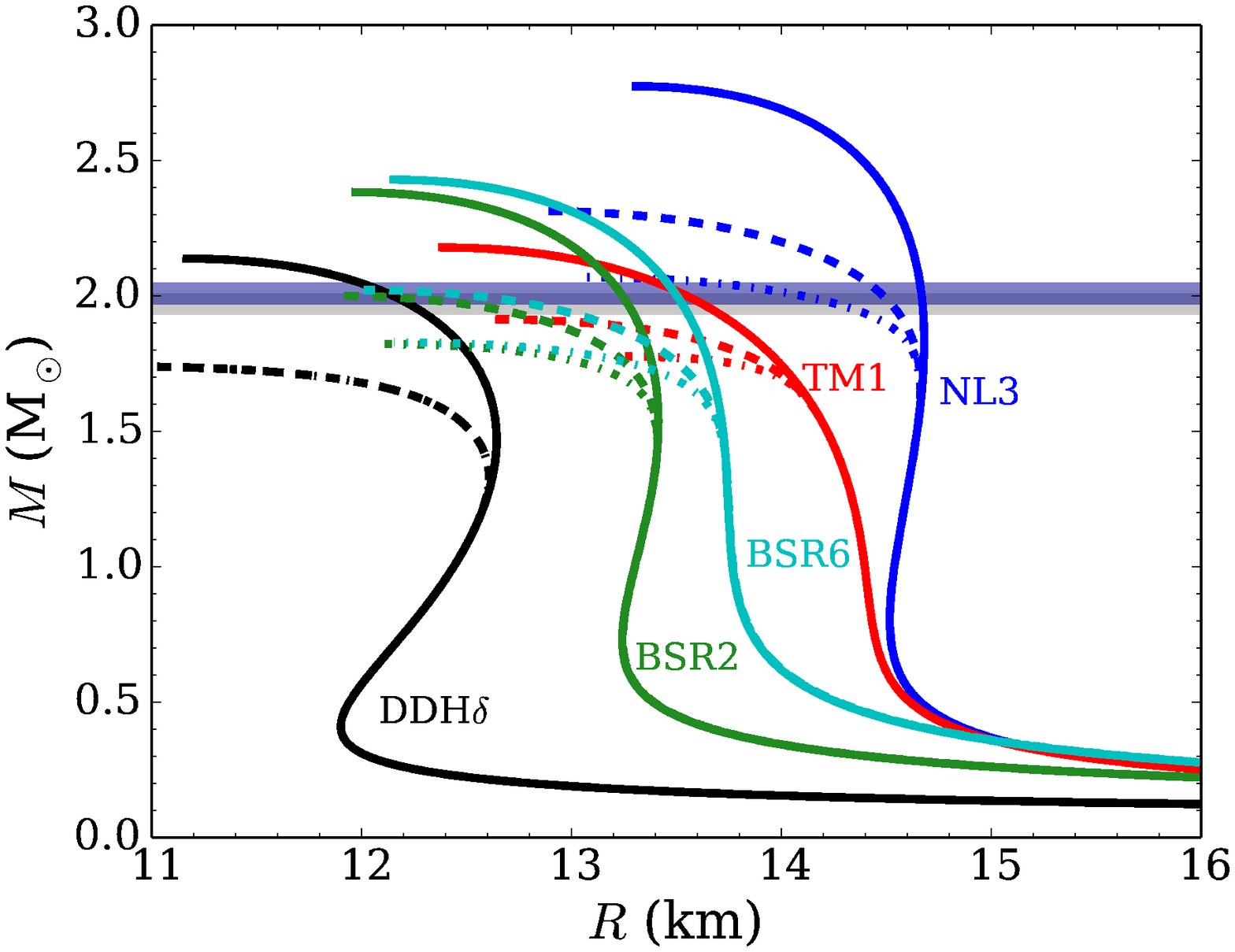}
%\end{tabular}
\caption{Mass-radius relation for the various RMF models: {\bf noY} ,
  {\bf Y} and  {\bf Yss}. The horizontal lines indicate the
  constraints set by the pulsars  PSR J0348+0432 and PSR J1614-2230.}
\label{rmf:mr}
\end{figure}

Unified EOS are built for all the models and the TOV equations
solved. The mass-radius relations $M-R$ of all models, nucleonic and hyperonic,
are plotted in Fig.~\ref{rmf:mr}. The models were chosen such that  nucleonic EOS predict stars with a mass above  $2\;M_\odot$.
We can observe that the same occurs for all {\bf Y}
models, except for  the one built with  DDH$\delta$. With respect to
the {\bf Yss} models, only the ones built with NL3 or NL3$\omega\rho$
satisfy the constraint set by PSR J0348+0432. Although models have been
distributed between two figures so that they are not too crowded,  it
is still possible to see that the radius of a $1.5\, M_\odot$ star
varies between $\sim$ 12.6 and 14.6  km.  Another conclusion is that
the onset of hyperons occurs for a mass $\sim 1.4\;M_\odot$ or above,
except for the DDH$\delta$ model. 

% MR RELATIONS

Models with the largest values of $L$ only predict droplet-like
clusters in the inner crust, in accordance with the results of
\cite{oyamatsu07},
see Table~\ref{tab:rmf}.

 In Table~\ref{rmf:compo} we gather some of the properties of the
hyperonic stars, including the central baryonic density, the maximum  mass,
the onset density of each hyperon and the corresponding mass of the star. It is
seen that, for the choice of meson-hyperon couplings described above, the
first hyperon to set in in all models is the $\Lambda$-meson. The
second hyperon is the $\Xi^-$ hyperon, again in all models. This
hyperon is favored with respect to  $\Sigma^-$ because of the
attractive $\Xi$-potential in nuclear matter. The third meson to set
in, when it exists in stable NS, is either  $\Sigma^-$ or  $\Xi^0$. 
$\Sigma^-$ appears only when the $\sigma^*$-meson is not included in
the calculation, because the attractive effect of the $\sigma^*$-meson
is stronger for $\Xi$-hyperons due to their double strangeness charge.

\begin{table*}
\begin{ruledtabular}

\center\begin{tabular}{llcc|lcclcclcc|cc}
EOS& Model&$n_{\rm max}$&$M_{\rm max}^{\rm Y}$& Y$_1$&$n_{\rm Y_1}$&$M_{\rm Y_1}$& Y$_2$&$n_{\rm Y_2}$&$M_{\rm Y_2}$& Y$_3$&$n_{\rm Y_3}$&$M_{\rm Y_3}$& $n_{\rm DU}$&$M_{\rm DU}$\\
   &      & (fm$^{-3}$) & $(\msun)$           &      & (fm$^{-3}$) & $(\msun)$&         & (fm$^{-3}$) & $(\msun)$   &      & (fm$^{-3}$) & $(\msun)$&  (fm$^{-3}$) & $(\msun)$\\
\hline

NL3 & {\bf Yss}     &0.77 &2.07&  $\Lambda^0$ &0.28 & 1.52& $\Xi^-$ &0.33 & 1.75&$\Xi^0$ &0.57 & 2.03 & 0.20 & 0.84\\
   & {\bf Y}  &0.78 &2.31&  $\Lambda^0$ &0.28 & 1.52& $\Xi^-$ &0.35 & 1.85&-&-&-& 0.20 & 0.84\\

\hline

NL3${\omega \rho}$ & {\bf Yss} &0.80 &2.14 & $\Lambda^0$  &0.31 &1.59
&$\Xi^-$&0.34 & 1.74 & $\Xi^0$ & 0.65 &2.13  & 0.36&1.80\\
     & {\bf Y} & 0.79 & 2.34 & $\Lambda^0$ & 0.31 & 1.58 &$\Xi^-$ &  0.34
& 1.78 & $\Sigma^-$ & 0.49 &2.17 & 0.37& 1.89 \\
\hline

DDME2& {\bf Yss} &0.87 &1.84  &$\Lambda^0$ &0.35 &1.46 &  $\Xi^-$ &0.37
&1.53 &$\Xi^0$ & 0.72 & 1.82 & 0.42& 1.62\\
     & {\bf Y} &0.93 &2.09  &$\Lambda^0$ &0.35 &1.46 &  $\Xi^-$
     &0.37 &1.56 &$\Sigma^-$ & 0.41 & 1.68 & 0.50 & 1.86\\
\hline

GM1 & {\bf Yss} & 0.82& 1.85& $\Lambda^0$& 0.35 & 1.48& $\Xi^-$ &0.40 & 1.64& $\Xi^0$ &0.70 & 1.84 & 0.28 & 1.10\\
    & {\bf Y} &0.92 &1.99 & $\Lambda^0$ &0.35 & 1.48&  $\Xi^-$ &0.41 & 1.67&-&-&-& 0.28 & 1.10\\
\hline

TM1 & {\bf Yss} &0.73& 1.78& $\Lambda^0$ & 0.35 & 1.52& $\Xi^-$ &0.39 & 1.63&$\Xi^0$ &0.72 & 1.78 & 0.21 & 0.81\\
    & {\bf Y} &0.85 &1.92 & $\Lambda^0$ &0.32 & 1.40&  $\Xi^-$ &0.42 & 1.70&-&-&-& 0.21 & 0.81\\
\hline

DDH$\delta$ &{\bf Y}&1.46 &1.74  &$\Lambda^0$ & 0.34&1.15 &
$\Xi^-$&0.51 & 1.53 &$\Sigma^-$ & 0.58 & 1.59 & 0.57& 1.59\\
\hline

DD2 & {\bf Yss}& 0.89& 1.69 & $\Lambda^0$ & 0.34&1.32&  $\Xi^-$& 0.37&
1.44& $\Xi^0$ & 0.73 & 1.68 & 0.42 & 1.52\\
  & {\bf Y}& 1.02& 2.00 & $\Lambda^0$ & 0.34&1.32& $\Xi^-$& 0.37&
  1.45 &$\Sigma^-$&0.41& 1.57& 0.45 & 1.66\\

\hline
BSR2 & {\bf Yss} & 0.84 & 1.84 & $\Lambda^0$ & 0.34  &1.37 &  $\Xi^-$&
0.39& 1.54 & - & - & - & 0.39& 1.54\\
     & {\bf Y} & 0.89 & 2.00 & $\Lambda^0$ & 0.34  &1.38 &  $\Xi^-$&
0.42& 1.65 & $\Sigma^-$ & 0.51 & 1.81 & 0.39 & 1.58\\

\hline
BSR6 & {\bf Yss} & 0.84 & 1.84 & $\Lambda^0$ & 0.33  &1.34 &  $\Xi^-$&
0.38& 1.54 & $\Xi^0$ & 0.81 & 1.83& 0.27 & 1.00\\
       & {\bf Y} & 0.87 & 2.03 & $\Lambda^0$ & 0.33  &1.36 &  $\Xi^-$&
0.42& 1.67 & $\Sigma^-$ & 0.57 & 1.91 & 0.27 & 1.00\\

\end{tabular}
\caption{Properties of hyperonic RMF models. For given EOS and hyperonic model,
the central density $n_{\rm max}$ (in fm$^{-3}$) at the
maximum mass $M_{\rm max}$ (in $M_\odot$) is given. The next columns
list the type of hyperons ${\rm Y_i}$ that appear with increasing density,
the density above which they do: $n_{\rm Y_i}$ (in fm$^{-3}$), and the
corresponding mass: $M_{\rm Y_i}$ (in $M_\odot$). The last two columns indicate the density and 
mass threshold,  $n_{\rm DU}$ and $M_{\rm DU}$ respectively above which the nucleonic DUrca process to operate.}
\label{rmf:compo}
\end{ruledtabular}
\end{table*}

\subsection{Non-relativistic unified EOS}

To construct non-relativistic unified equations of state we proceed as follows.
We select a large set of different Skyrme functionals proposed in the recent nuclear physics literature. At low density we variationally determine the nucleus $A$ and $Z$ number, as well as the volume $V_{WS}$ of the Wigner-Seitz cell and the density of the free neutron component $n_{\rm g}$ after neutron drip, employing the same Skyrme functional for both the nucleus and the free neutrons
 \cite{WSpaper}. The baryonic part of the Wigner-Seitz cell energy is written as  
\begin{equation}  
E_{\rm WS}(A,Z,n_{\rm g},V_{\rm WS})=V_{\rm WS}{\cal E}_{\rm Sky} 
+ E^{\rm vac} +\delta E \label{energy}.
\end{equation}
Here, ${\cal E}_{\rm Sky}(n_{\rm g})$ is the energy density of homogeneous neutron matter as given from the chosen Skyrme functional, $E^{\rm vac}(A,Z)$ is the vacuum energy of a nucleus of mass number $A$ and charge $Z$, and the extra term 
$\delta E=\delta E_{\rm bulk}+\delta E_{\rm surf}+\delta E_{\rm Coul}$ 
corresponds to the bulk, surface and Coulomb in-medium modifications.

\begin{figure}
\includegraphics[width=0.78\linewidth]{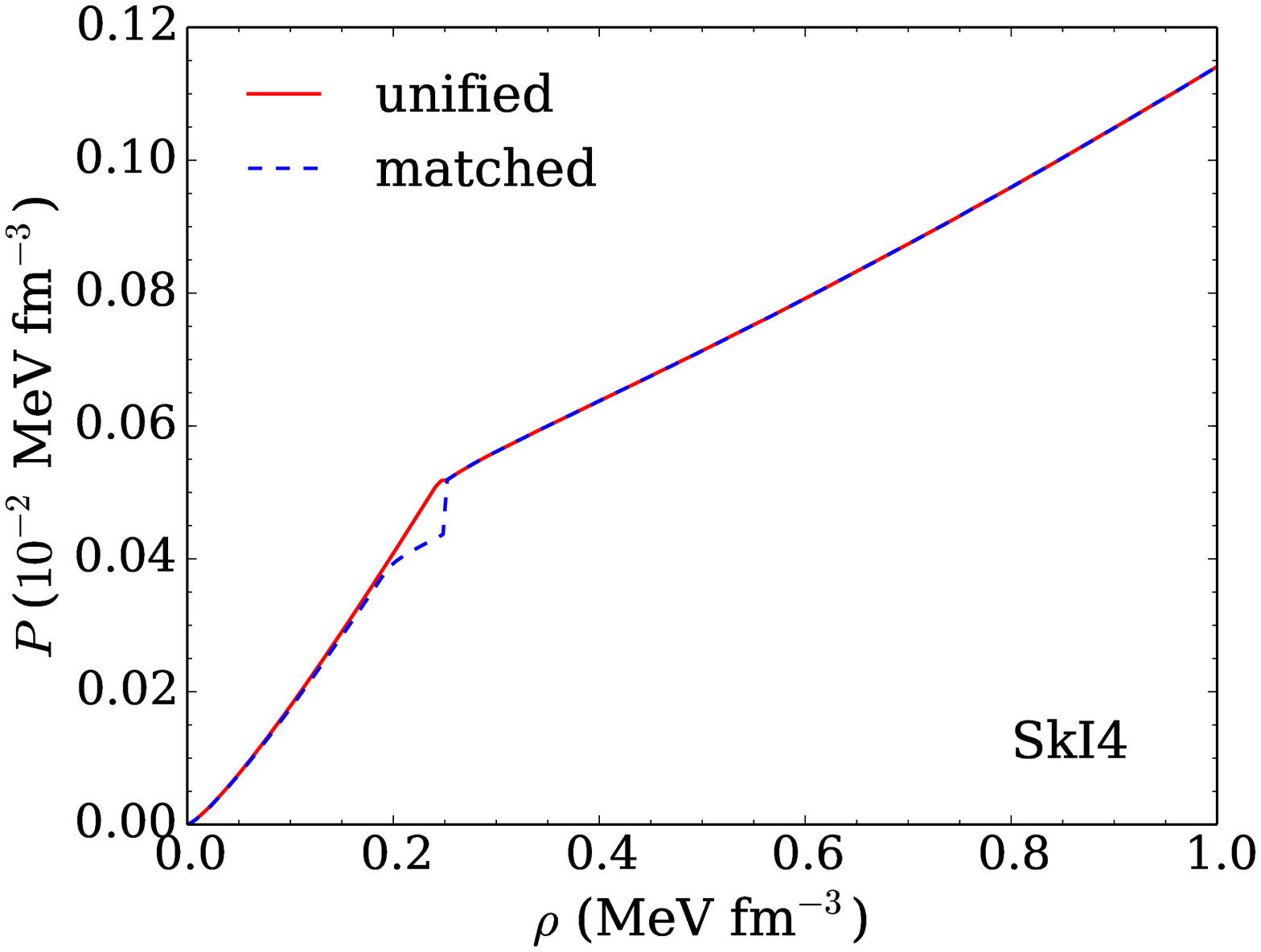}
\includegraphics[width=0.78\linewidth]{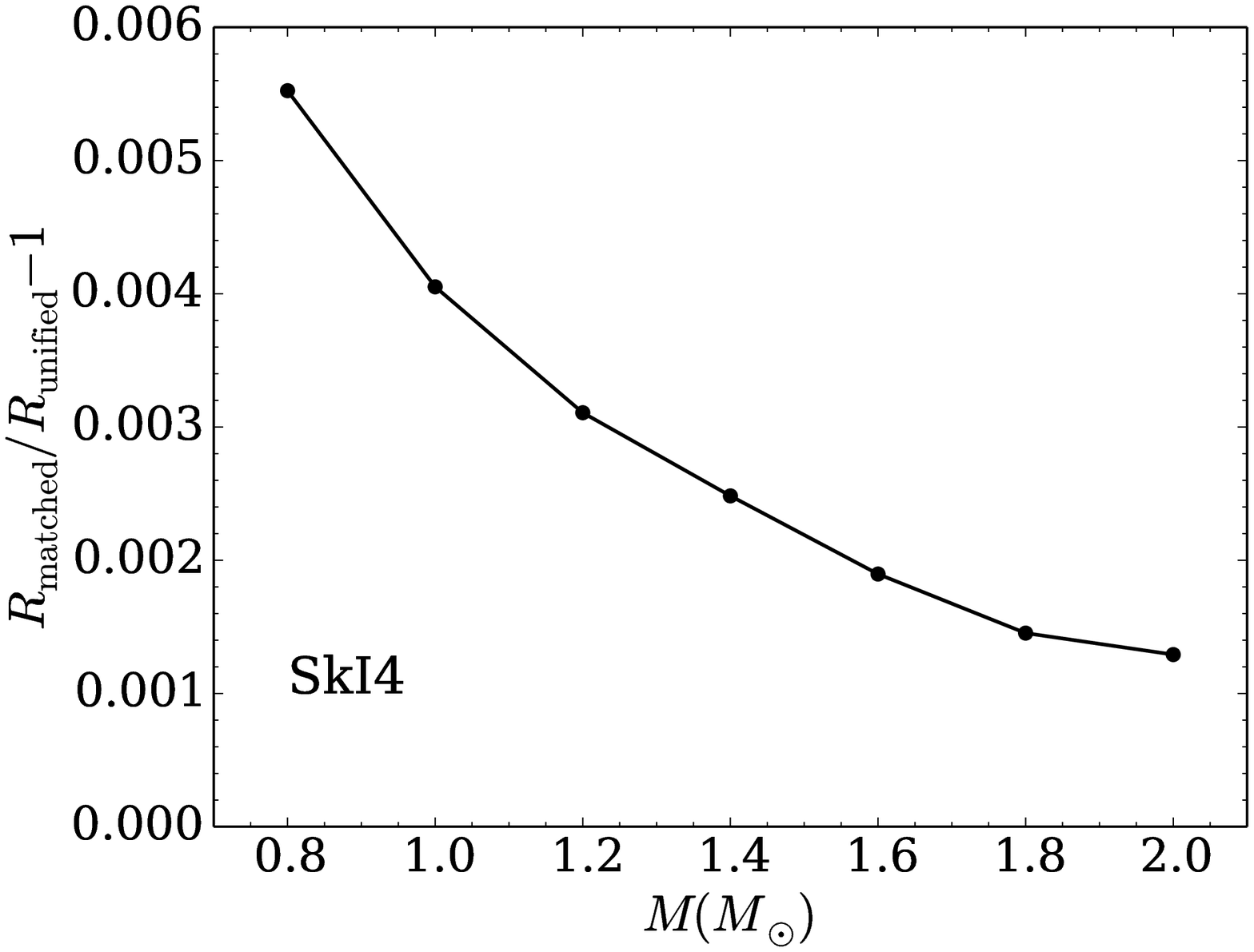}
\caption{(n,p,e) matter in $\beta-$equilibrium.
  Upper panel: total pressure versus
  total energy density for the unified EOS (solid lines) 
  and using experimental masses when available (dashed lines). 
  Lower panel: relative deviation of the NS radius as a function of mass 
  with the two prescriptions shown on the top part. 
  The SKI4 functional is used.   
}
\label{matching}
\end{figure}

For the vacuum energy,  we employ a  compressible liquid-drop (CLDM) parameterization \cite{Danielewicz2009}.
The coefficients of this mass formula are fitted out of 
Hartree-Fock calculations in slab geometry, using the same 
Skyrme effective interaction which is employed for the free neutron component. 
The absence of shell effects and curvature terms in this analytic parametrization implies 
that a mass shift which, depending on the Skyrme interaction, can be as high as 0.5 MeV/nucleon, 
is observed with respect to experimentally measured masses.
As a consequence, the EOS of  the external part of the outer crust
differs from the one we would get employing experimental data. This is true even for recent 
sophisticated Skyrme functional which have shown, if full HFB calculations are performed, 
a very good agreement with experimentally measured nuclear masses
\cite{BSk22-26}. 
An example is given in Fig.~\ref{matching}, which shows for a representative Skyrme functional 
the discontinuity in the baryonic pressure obtained if an EOS using 
experimental masses is matched with the unified prescription.   

% Matching of the outer crust

However,  the deviation between the CLDM-based EOS and the one obtained when the experimental mass data are used is
small enough to impact the $M(R)$ relation to less than 1\%, as one can see in the right panel of Fig.~\ref{matching}.

The bulk in-medium correction to the nuclear energy $\delta E_{\rm bulk}$
is approximated by 
\begin{equation}  
\delta E_{\rm bulk}(A,Z,n_{\rm g})=-\frac{A}{n_{\rm s}}{\cal E}_{\rm Sky},
\end{equation}
where $A/n_{\rm s}(\delta)$ represents the equivalent cluster volume corresponding to the given  isospin asymmetry $\delta$, evaluated in the nuclear bulk. For a nucleus in the vacuum we take 
for the bulk asymmetry the estimation from the droplet model~\cite{ldm}:
 
\begin{equation}
\delta=\delta_0 = \left[{(1-2{Z\over A})+ \frac{3a_c}{8Q} \frac{Z^2}{A^{5/3}}}\right]
/\left({1+ \frac{9 J}{4QA^{1/3}}}\right).
\label{eq_asym_deltar}
\end{equation}
In this equation, $J$ is the symmetry energy per nucleon at the saturation
density of symmetric matter,  $Q$ is the surface stiffness coefficient, and $a_{\rm c}$ is the Coulomb parameter.
In the presence of an external neutron gas of density $n_{\rm g}$,  
the bulk asymmetry defined by Eq.~(\ref{eq_asym_deltar}) 
is generalized such as to account for the contribution of the gas as~\cite{Panagiota2013}: 
 
\begin{eqnarray}
\delta(A,Z,n_{\rm g})&=&\left( 1-\frac{n_{\rm g}}{n_{\rm s}(\delta)}\right) \delta_0
+\frac{n_{\rm g}}{n_{\rm s}(\delta)},
\label{paperpana1:eq:deltacl}
\end{eqnarray}
where $\delta_0$ is the asymmetry value given by Eq.~(\ref{eq_asym_deltar}) 
considering only the bound part of the cluster. 
For details, see \cite{Panagiota2013,esym_paper,WSpaper}.

The Coulomb energy shift $\delta E_{\rm Coul}$ is due to the screening effect of the electrons,  and it is evaluated in the standard Wigner-Seitz approximation \cite{BBP}. 
The residual energy shift corresponds to the in-medium modification of the surface tension in  the inner crust. This term can be evaluated in the Extended-Thomas-Fermi approximation  \cite{douchin,esym_paper,Aymard2014}. 
For the applications of the present paper,
this correction has been neglected. The error induced by this approximation
on the $M(R)$ relation is quantified below in this section, and shown to be 
reasonably small. However, this effect, together with
the curvature terms which are also neglected, is important 
for a precise determination of the transition density.
For this reason we leave the study of the functional dependence of the 
transition density to future work.

Since the droplet phase is known to be the dominant pasta phase in $\beta$ equilibrium 
\cite{pasta_RMF}, we have not considered possible deviation from spherical symmetry in 
the nucleus functional. 

The Wigner-Seitz energy density from Eq.~(\ref{energy}) is minimized with respect to its arguments 
with the additional requirement of $\beta$ equilibrium, thus leading to the equilibrium composition 
of the neutron star crust at each baryonic density value \cite{WSpaper}. 
 In the absence of deformation degrees of freedom the crust-core transition occurs via a narrow phase coexistence domain \cite{WSpaper}. The transition to the core is defined by the high density
border of this first order phase transition region.
 
As stated above, a precise treatment of this transition requires proton shell effects, 
curvature terms, modifications of the surface tension, and deformation degrees of freedom.
However the energy density landscape turns out to be extremely flat close to the transition point 
\cite{douchin,WSpaper}. This means that the approximations employed in Eq.~(\ref{energy}) 
prevent a precise determination of the transition density, but do not affect the 
density behavior of the pressure.

\begin{figure}
\includegraphics[width=0.8\linewidth]{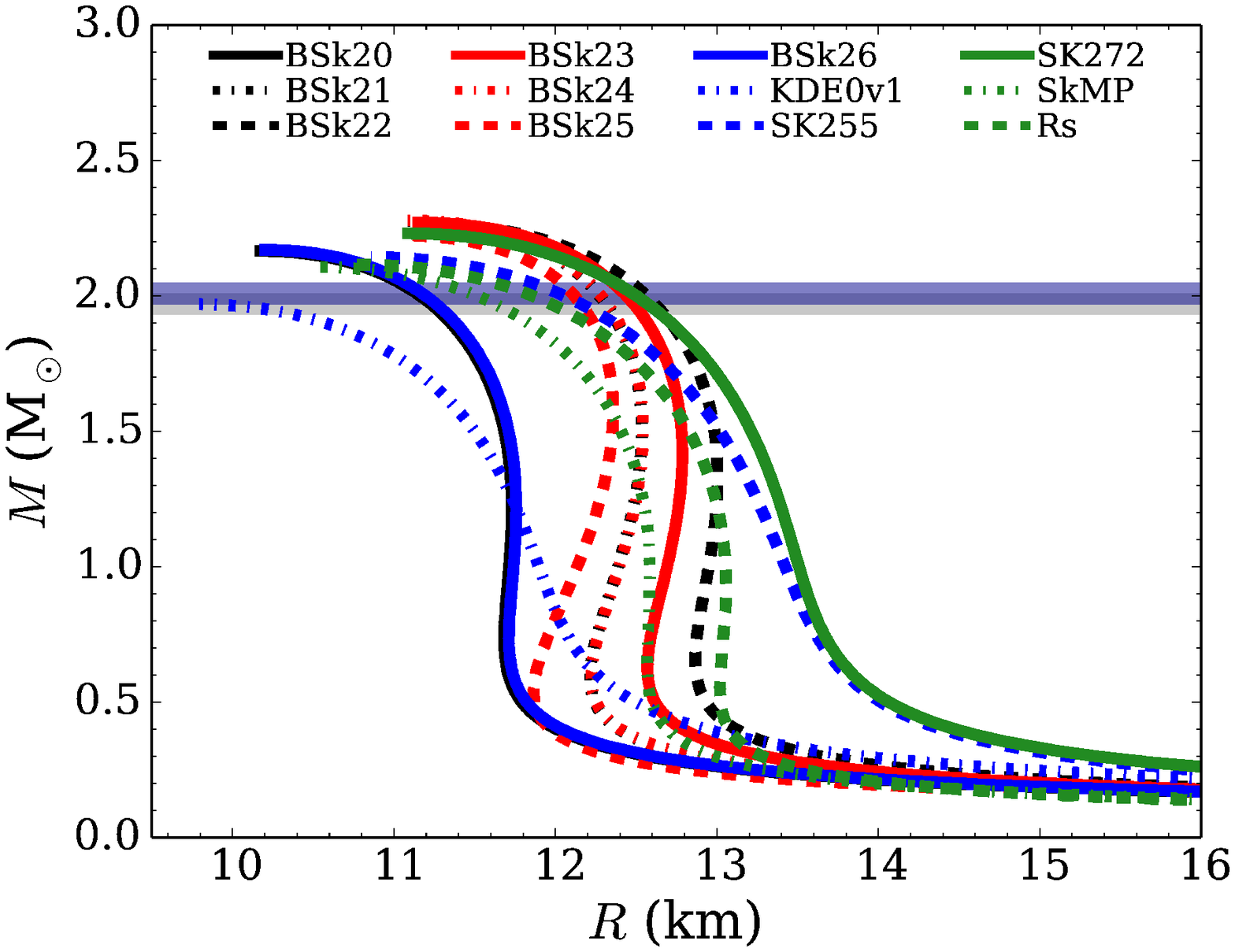}
\includegraphics[width=0.8\linewidth]{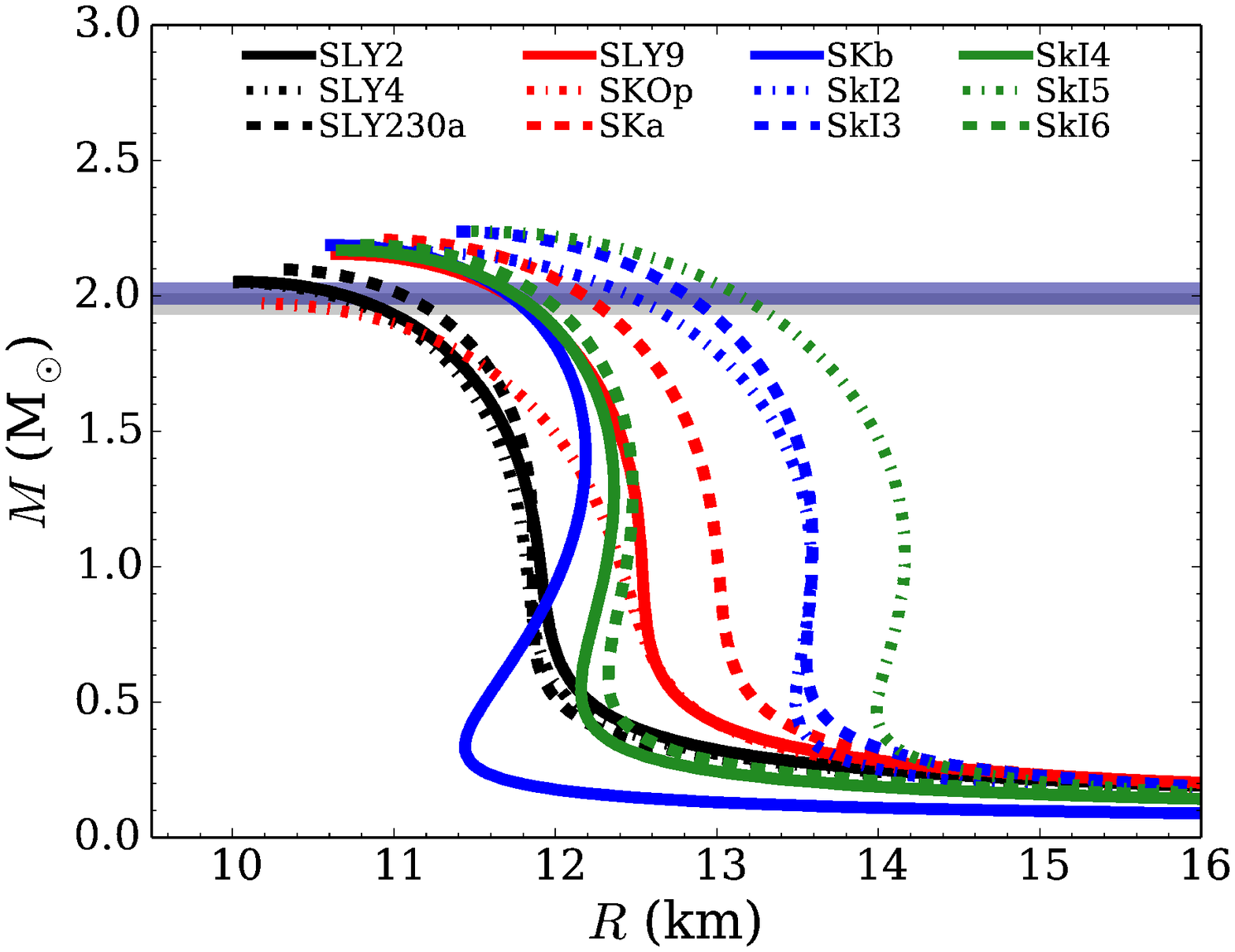}
\caption{Mass-radius relations for Skyrme models.}
\label{sk:mr}
\end{figure}

The $M(R)$  relation is then obtained integrating the TOV equations. 
Only the functionals which produce, without hyperonic degrees of freedom, maximum 
NS masses of at least, within $1\%$ accuracy, 
$2\,M_{\odot}$, and which are causal up to the highest densities met in such massive
stars are kept for the following analysis. The list of the functionals which have been retained, 
and the corresponding EOS parameters, are listed in Table  \ref{table:NNparam}. With the exception of BSk20
 and BSk26 models, the causality condition actually holds also up to the maximum mass.
The resulting $M(R)$ relation is shown in Fig.~\ref{sk:mr}. It is seen
that $1.0\,M_\odot - 1.5\, M_\odot$ stars span a radius range $\sim 3$ km wide, from  $\sim 11.5$ km  to $\sim 14.2$ km. 

\begin{center}
\begin{table*}
\begin{ruledtabular}
\begin{tabular}{lllllllllccl}
%\hline
functional & $n_{\rm s}$   & $E_{\rm s}$ & $K$      & $J$   & $L$   & $K_{\rm sym}$ & $M_{\rm max}^{\rm noY}$ & ${\rm v^2_{\rm sound}}(2\,M_{\odot})$ &$ n_{\rm DU}$& $ M_{\rm DU}$& Ref.\\
$~$        & (fm$^{-3}$) & (MeV) & (MeV)    & (MeV) & (MeV) & (MeV)         & ($M_{\odot}$)   & (c$^2$)                   &(fm$^{-3}$) & $(\msun)$   & $~$\\
\noalign{\smallskip}\hline
SKa          & 0.155      & -15.99 & 263.16 & 32.91 & 74.62 & -78.46  & 2.22 & 0.61 &0.37& 1.23 & \cite{SKab}\\
SKb          & 0.155      & -16.00 & 263.0  & 33.88 & 47.6  &  -78.5  & 2.20 & 0.63 &1.67   & -    & \cite{SKab}\\
SkI2         & 0.1575     & -15.77 & 241.0  & 33.4  & 104.3 &   70.7  & 2.17 & 0.56 &0.26& 0.92 & \cite{SkI2-5} \\
SkI3         & 0.1577     & -15.98 & 258.2  & 34.83 & 100.5 &   73.0  & 2.25 & 0.54 &0.26& 0.92 & \cite{SkI2-5}\\
SkI4         & 0.160      & -15.95 & 247.95 & 29.50 & 60.39 & -40.56  & 2.18 & 0.64 &0.52& 1.64 & \cite{SkI2-5}\\
SKI5         & 0.156      & -15.85 & 255.8  & 36.64 & 129.3 &  159.5  & 2.25 & 0.51 &0.22& 0.86 & \cite{SkI2-5}\\
SkI6         & 0.159      & -15.89 & 248.17 & 29.90 & 59.24 & -46.77  & 2.20 & 0.62 &0.51& 1.66 & \cite{SkI6}\\
Sly2         & 0.161      & -15.99 & 229.92 & 32.00 & 47.46 & -115.13 & 2.06 & 0.78 &1.22& - & \cite{SLY2and9}\\
Sly230a      & 0.160      & -15.99 & 229.90 & 31.99 & 44.30 & -98.3   & 2.11 & 0.72 &0.82& 2.00 & \cite{SLY230a}\\
Sly4         & 0.159      & -15.97 & 230.0  & 32.04 & 46.00 & -119.8  & 2.06 & 0.79 &1.14   &-     & \cite{SLY4}\\
Sly9         & 0.151      & -15.80 & 229.84 & 31.98 & 54.86 & -81.42  & 2.16 & 0.65 &0.56& 1.72 & \cite{SLY2and9}\\
SkMP         & 0.157      & -15.56 & 230.87 & 29.89 & 70.31 & -49.82  & 2.11 & 0.66 &0.43& 1.32 & \cite{SkMP}\\
SKOp         & 0.160      & -15.75 & 222.36 & 31.95 & 68.94 & -78.82  & 1.98 & 0.55 &0.58& 1.53 & \cite{SKOp}\\
KDE0V1       & 0.165      & -16.23 & 227.54 & 34.58 & 54.69 & -127.12 & 1.98 & 0.57 &1.79   &-     & \cite{KDE0v} \\
SK255        & 0.157      & -16.33 & 254.96 & 37.4  & 95.0  &  -58.3  & 2.15 & 0.61 &0.25& 0.76 & \cite{SK255and272} \\  
SK272        & 0.155      & -16.28 & 271.55 & 37.4  & 91.7  &  -67.8  & 2.24 & 0.59 &0.26& 0.80 & \cite{SK255and272} \\
Rs           & 0.157      & -15.53 & 236.7  & 30.58 & 85.7  &  -9.1   & 2.12 & 0.62 &0.32& 1.06 & \cite{Rs} \\
BSk20        & 0.1596     & -16.080& 241.4  & 30.0  & 37.4  & -136.5  & 2.17 & 0.77 &1.49& - & \cite{BSk20-21}\\ 
BSk21        & 0.1582     & -16.053& 245.8  & 30.0  & 46.6  &  -37.2  & 2.29 & 0.60 &0.45& 1.60 & \cite{BSk20-21}\\ 
BSk22        & 0.1578     & -16.088& 245.9  & 32.0  & 68.5  &   13.0  & 2.27 & 0.56 &0.33& 1.15 & \cite{BSk22-26}\\
BSk23        & 0.1578     & -16.068& 245.7  & 31.0  & 57.8  &  -11.3  & 2.28 & 0.58 &0.38& 1.34 & \cite{BSk22-26}\\
BSk24        & 0.1578     & -16.048& 245.5  & 30.0  & 46.4  &  -37.6  & 2.29 & 0.60 &0.45& 1.60 & \cite{BSk22-26}\\
BSk25        & 0.1587     & -16.032& 236.0  & 29.0  & 36.9  &  -28.5  & 2.23 & 0.58 &0.47& 1.63 & \cite{BSk22-26}\\
BSk26        & 0.1589     & -16.064& 240.8  & 30.0  & 37.5  & -135.6  & 2.18 & 0.76 &1.46& - & \cite{BSk22-26}\\
\end{tabular}
\caption{Nuclear and astrophysical properties of considered Skyrme functionals. Saturation density $n_{\rm s}$, energy per nucleon $E_{\rm s}$, incompressibility (at saturation)
$K$, symmetry energy (at saturation) $J$, symmetry energy slope $L$, and symmetry incompressibility $K_{\rm sym}$, all calculated at saturation point. ${\rm M}_{\rm max}^{\rm noY}$ is the maximum mass and $v^2_{\rm sound}(2\,M_{\odot})$ is the square of the sound 
speed at the center of a NS with  $2\;M_\odot$. $n_{\rm DU}$ and ${\rm M}_{\rm DU}$ are respectively the density and 
mass threshold above which the nucleonic DUrca is switched on.}
\label{table:NNparam}
\end{ruledtabular}
\end{table*}
\end{center}

% MR RELATIONS

\begin{figure}
\includegraphics[width=0.8\linewidth]{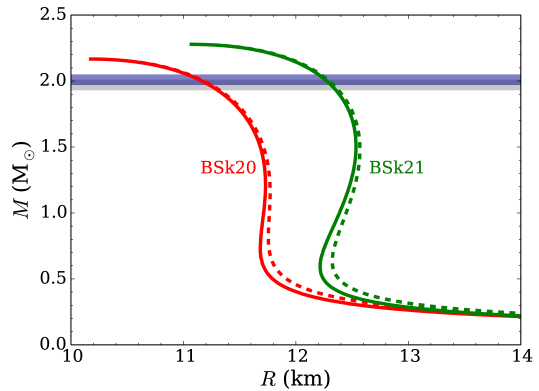}
\caption{Mass as a function of the radius for the BSk20 and BSk21 functionals. Full lines: full microscopic HFB calculation from \cite{BSk20-21}. Dashed lines: our model for the unified EOS.}
\label{MR_compa}
\end{figure}

In the case of some of the Skyrme functionals developed by the Brussels group
\cite{BSk20-21}, the $M(R)$ relation has already been calculated with a unified EOS obtained by numerically solving in the crust the full Hartree-Fock-Bogoliubov problem in the Wigner-Seitz cell. A comparison with our results using the simplified CLDM allows quantifying the error which is made because of the different approximations employed to get an analytic model, namely the lack of shell effects, curvature terms, and in-medium modifications of the surface tension. This comparison is shown
in Fig.~\ref{MR_compa}. We can see that the estimation of the mass is never affected by the approximations (the dashed and full lines are very similar for $M\geq 1\,M_{\odot}$ on the right part of the figure), while for a fixed mass a deviation is observed in the radius, deviation which increases as expected with decreasing mass. We consider that this comparison is representative of the systematic error which is made for all functionals due to the limitations of the model. We have checked that this mass dependent error bar is always smaller than the size of the symbols and width of the lines of all the figures of the present paper.

% MR  COMPA

\section{Results}

In the present section we discuss the uncertainties on the
determination of the radius and the crust thickness of a star
associated with the
 models presented in the previous section. 
In the next subsection we will also discuss how the radius and the crust thickness of
NS are related with two properties at saturation: the
incompressibility and the symmetry energy slope.
We will then proceed to impose a set of terrestrial constraints and select  the models
that satisfy all constraints, or miss utmost two by less than 10\%,
and will discuss how the uncertainties on the
determination of the radius and the crust thickness of a star
previously obtained are affected. The final subsection focuses on the DUrca process and 
the possible constraint that could be put on $L$, the NS radius or the EOS thanks to the astrophysical
 observations of thermal states of NS.

\subsection{Radius and crust thickness}
\label{sec:rad}

%RADIUS VS. L AND K FOR DIFFERENT MASSES %%%%%%%%%%%%%%%%%%%%%%%%%%%%%%%%%%%%%%%%%%%%%%%%%%%%%%%%%%%%%%%%%%%%%%%%%%%%%%%%%%%%%%%%%
\begin{figure*}
\resizebox{\hsize}{!}{\includegraphics[width=0.8\columnwidth]{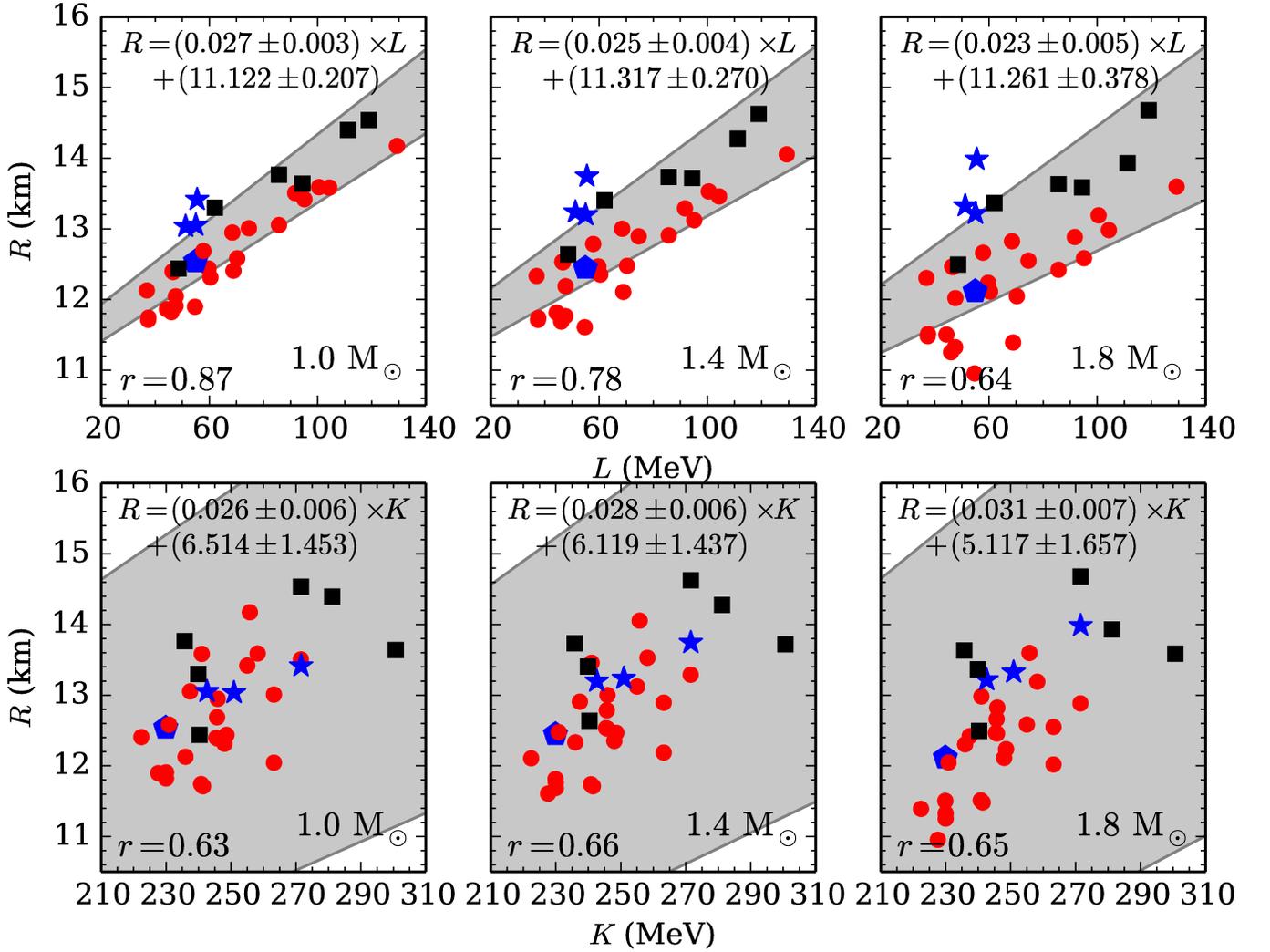}}
\caption{Radii of purely nucleonic NS as a function of the symmetry energy slope $L$ (upper plots) and the incompressibility of symmetric matter $K$ (lower plots) %and central density 
for different masses (1.0, 1.4 and 1.8 $\msun$ from left to right). The red dots indicate Skyrme models, the black ones RMF models. The blue symbols (stars for RMF and a pentagon for Sly9)  correspond to models which are at the intersection of all nuclear constraints in the $L-J$ plane: see Fig.~\ref{L-J}.
The shaded areas indicate the result of a linear regression, including the error bars in the fitted parameters. The correlation coefficient $r$ is indicated in each plot.
}
\label{rad}
\end{figure*}

In   Fig.~\ref{rad}, the radii of $1.0\,M_\odot$, $1.4\,M_\odot$ and $1.8\, M_\odot$ NS for a purely nucleonic core are plotted as a
function of the slope $L$ and the incompressibility $K$. 
We can see that the radii for the various EOS differ at most by $2.8\;{\rm km},
3.0\;{\rm km}$ and 3.7 km
for masses of $1.0\;M_\odot$, $1.4\;M_\odot$ and $1.8\,M_\odot$, respectively.
The uncertainty on the radius is connected with the nuclear
 properties of the EOS of the models used \cite{lattimer07}. In the
 next subsection we will restrict ourselves to the models that also
 satisfy other constraints both from  experiments and from theoretical
 calculations of pure neutron matter and will discuss how much this
 uncertainty changes.

We can also see from  Fig.~\ref{rad} that the radius appears well correlated with the slope of the symmetry energy $L$, especially for low mass stars. This correlation is still present for the more massive stars but the dispersion increases with the mass, as expected. Indeed,
 in \cite{carriere03} the authors have shown that the radius of  low
mass stars is well correlated with the neutron skin thickness of
$^{208}$Pb.  On the other hand, it has been discussed in
\cite{brown00,vinas09} that  the
neutron skin thickness is very sensitive to the 
density dependence of the nuclear symmetry energy and, in particular, to the slope parameter $L$ at the normal nuclear
saturation density. 
The correlation obtained in  \cite{carriere03}  corresponds,
therefore, to a
correlation between the star radius and the slope $L$. The authors take a set of four different models and within each span a wide range of neutron skin thicknesses by changing the density dependence of the symmetry
energy. The correlation between the star radius and the  neutron skin thickness of $^{208}$Pb is particularly strong for stars with masses $0.5\;M_\odot$
and $0.75\;M_\odot$. For $M=1.4\; M_\odot$, although a clear correlation is
still present, the spread of the distribution is wider showing a
larger model dependence. The
behavior was attributed to the stellar matter densities that were being
explored within each type of star: for low mass the main contribution
comes from densities close to the saturation density where all
models behave similarly because most of them are fitted to finite
nucleus properties. The properties of stars with larger masses are
also determined by
the high density EOS, corresponding to a range of densities where the higher order coefficients in the density expansion of the energy functional play an increasing role. 

Looking now
at the radius as a function of the incompressibility, a linear
correlation is also observed as indicated by the non-zero value of the correlation coefficient. However, the  spread of the
data for the three masses considered is considerably larger than when considering correlations between $L$ and the radii. 
This can be quantified by looking at the result of a fit, using a linear regression,
 of the radius $R$ for different masses,
 with a linear function $ax+b$, where $x=L$ or $K$. The result of the fit, including
the error bar on the two fitting parameters, is represented in Fig.~\ref{rad} as a shaded area.
In the case of $R(L)$ (upper panel), a well defined linear behavior can be extracted, even if the importance of higher order terms in the density expansion \cite{margueron15} can be inferred by the larger dispersion at high mass. 
On the contrary, the error in the $b$ parameter is so large that no
relevant information on $K$ can be extracted from the radius.
This indicates that, as far as isoscalar properties are concerned, 
the influence of higher order terms cannot be neglected.
An analytic parametrization for radii of NS with different masses 
in terms of properties of 
symmetric saturated matter was first discussed in \cite{lattimer07} and 
a quite complex dependence on $K$, 
skewness parameter $K'=27 n_{\rm s}^3 \left(\partial^3 E_{_{\rm NM}}/\partial n^3\right)_{n_{\rm s},\delta=0}$ and 
$L$  was highlighted.
 
\begin{figure}
%\begin{tabular}{cc}
\includegraphics[width=0.8\linewidth]{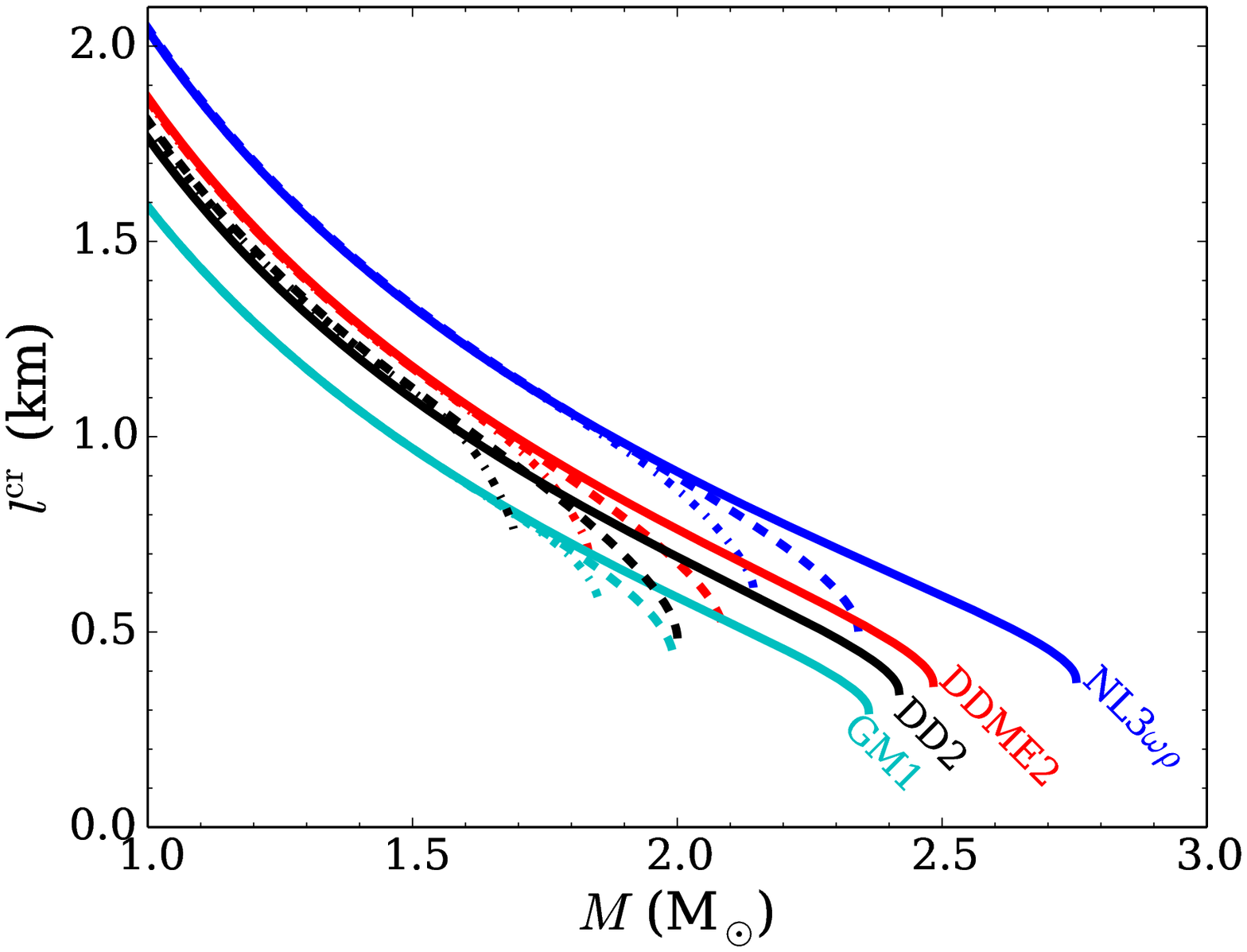}
\includegraphics[width=0.8\linewidth]{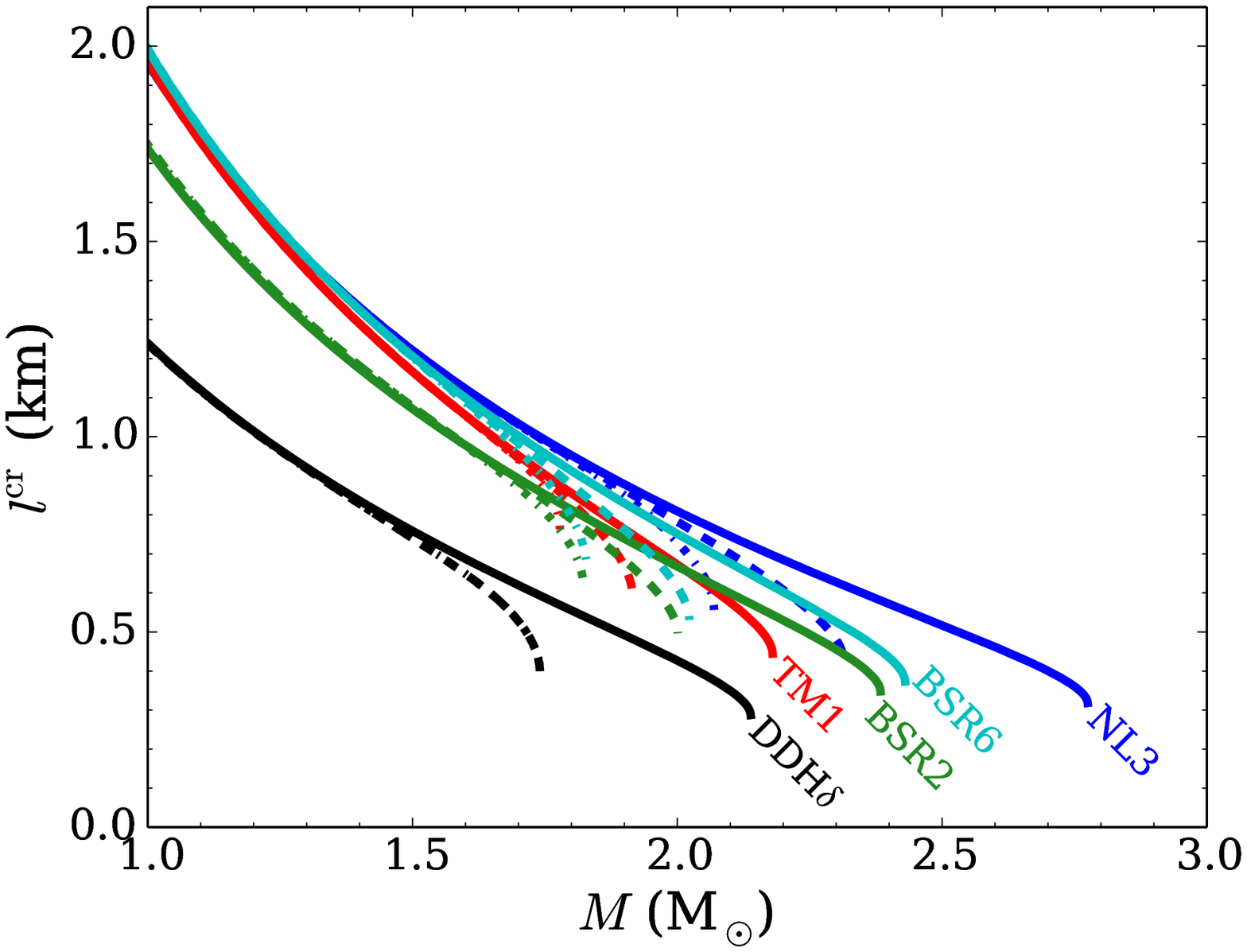}
%\end{tabular}
\caption{Mass vs. crust thickness relation for the various RMF models: {\bf noY} ,
  {\bf Y} and  {\bf Yss}. Line styles correspond to the ones used in Fig.~\ref{rmf:mr}.}
\label{crust}
\end{figure}

The crust thickness for the RMF models is plotted as a function of the star mass in Fig.~\ref{crust}. We do not show results for the Skyrme
parametrizations because the method used to describe the crust in
these models does
not allow for a precise calculation of the crust-core transition density as
explained before.
 For the models represented in Fig.~\ref{crust},  no correlations were found between the crust thickness and the
slope $L$ or  the incompressibility $K$ for stars with masses $1.0\,M_\odot$, $1.4\,M_\odot$,  and $1.8\,M_\odot$.  Excluding the DDH$\delta$model that predicts the
smallest thickness, we have obtained: $1.6\;{\rm km}<l^{\rm cr}<2.1$ km for a star with
$M= 1.0\;M_\odot$, $1.1\;{\rm km}<l^{\rm cr}<1.5$ km for
$1.4\;M_\odot$ and $0.7\;{\rm km}<l^{\rm cr}<1.1$ km for
$1.8\;M_\odot$.
For stars with masses 1.0$\;M_\odot$ and 1.4$\;M_\odot$  the upper
limits of the crust thickness are more than 30\% larger than the lower
limits. This difference raises to 50\% for the 1.8$\;M_\odot$ star.
%%%%%%%%%%%%%%%%%%%%%%%%%%%%%%%%%%%%%%%%%%%%%%%%%%%%%%%%%%%%%%%%%%%%%%%%%%%%%%%%%%%%%%%%%%%
%CRUST THICKNESS VS. L AND K FOR DIFFERENT MASSES %%%%%%%%%%%%%%%%%%%%%%%%%%%%%%%%%%%%%%%%%%%%%%%%%%%%%%%%%%%%%%%%%%%%%%%%%%%%%%%%%%%%%%%%%
% MR RELATIONS
%%%%%%%%%%%%%%%%%%%%%%%%%%%%%%%%%%%%%%%%%%%%%%%%%%%%%%%%%%%%%%%%%%%%%%%%%%%%%%%%%%%%%%%%%%%

\subsection{Comparison with nuclear constraints}

%PRESSURE FOR PNM AT N0 VS HEBELER AND GANDOLFI %%%%%%%%%%%%%%%%%%%%%%%%%%%%%%%%%%%%%%%%%%%%%%%%%%%%%%%%%%%%%%%%%%%%%%%%%%%%%%%%%%%%%%%%%
\begin{figure}
%\begin{tabular}{cc}
\includegraphics[width=0.8\linewidth]{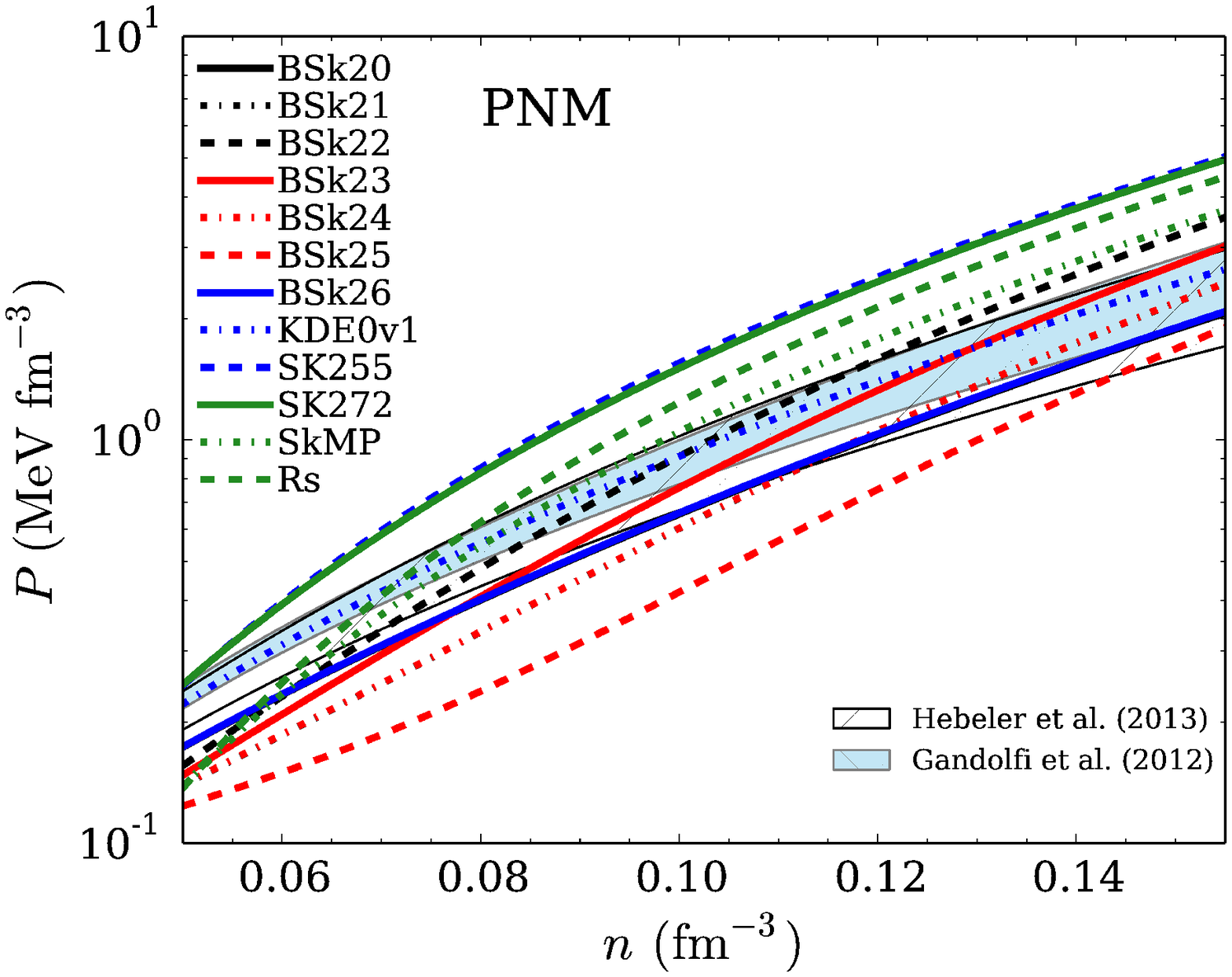}
\includegraphics[width=0.8\linewidth]{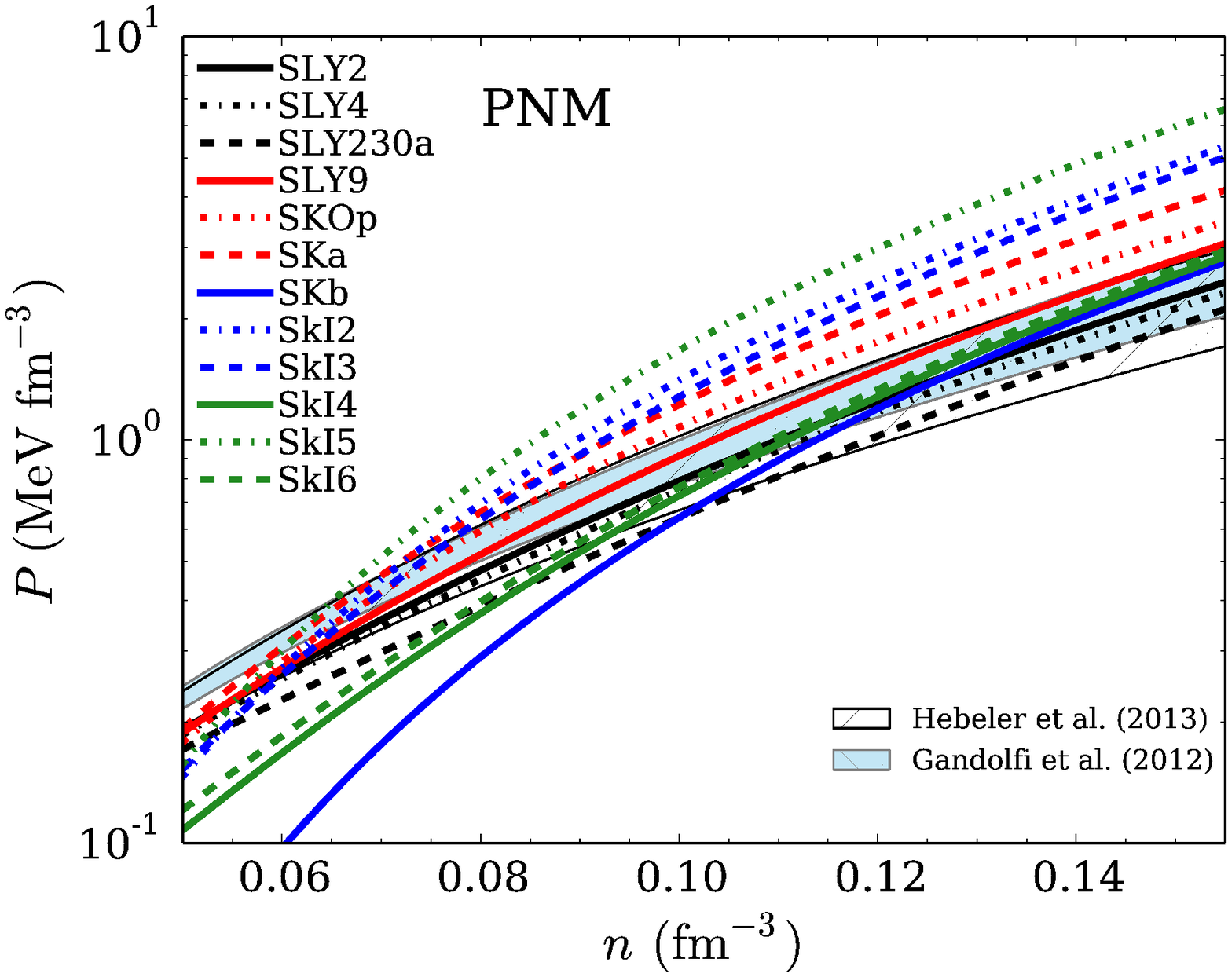}
%\end{tabular}
\caption{Pressure for pure neutron matter for Skyrme models and constraints by \cite{H13} and \cite{G12}.}
\label{pn0:sk}
\end{figure}

\begin{figure}
\includegraphics[width=0.8\linewidth]{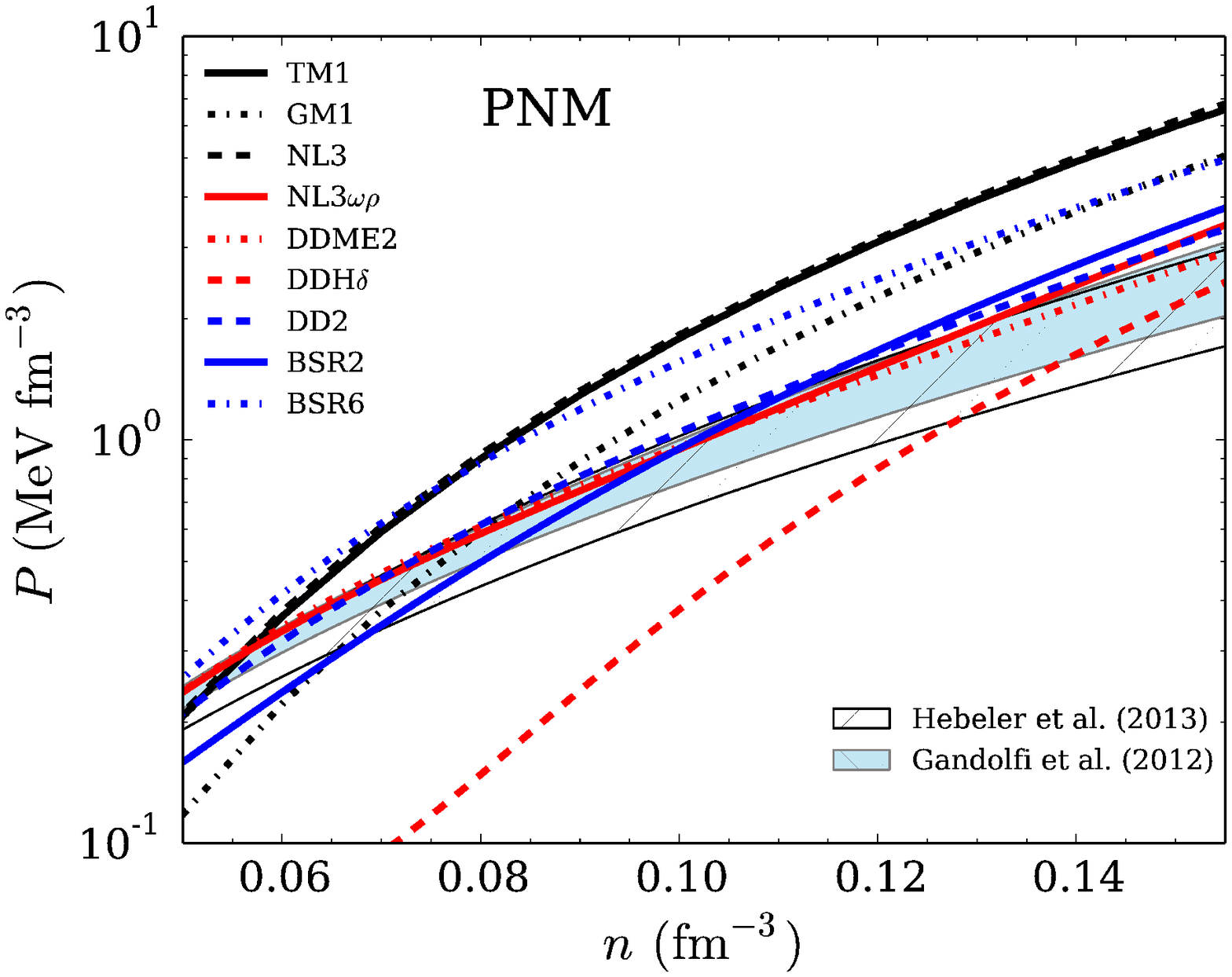}
\caption{Pressure for pure neutron matter for RMF models and constraints by \cite{H13} and \cite{G12}.}
\label{pn0:rmf}
\end{figure}
%%%%%%%%%%%%%%%%%%%%%%%%%%%%%%%%%%%%%%%%%%%%%%%%%%%%%%%%%%%%%%%%%%%%%%%%%%%%%%%%%%%%%%%%%%%

%L-J VS TSANG ET AL, 2012 %%%%%%%%%%%%%%%%%%%%%%%%%%%%%%%%%%%%%%%%%%%%%%%%%%%%%%%%%%%%%%%%%%%%%%%%%%%%%%%%%%%%%%%%%
\begin{figure}
\resizebox{\hsize}{!}{\includegraphics{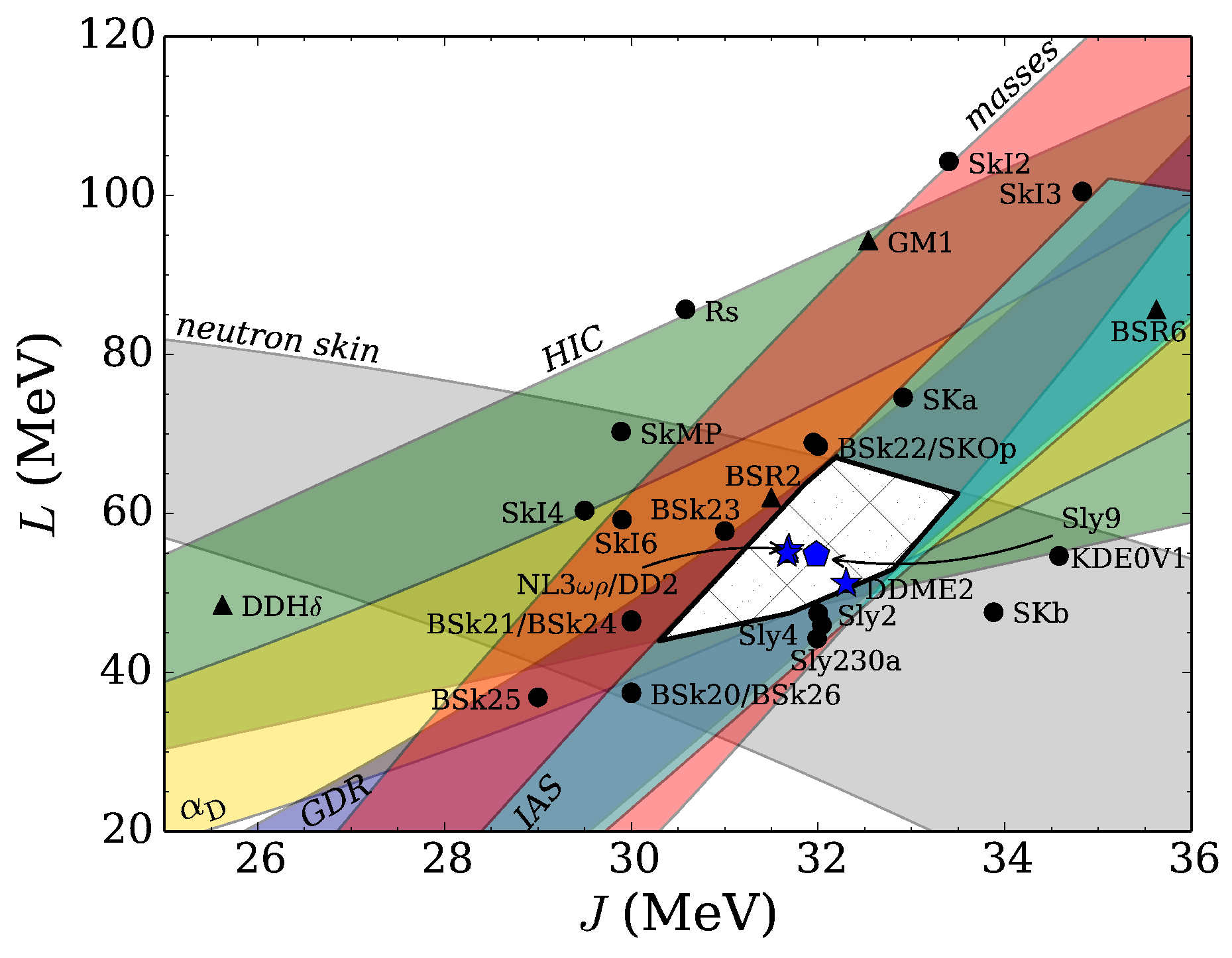}}
\caption{$L$ and $J$ parameters of all our EOS compared to various nuclear constraints (see text for details). The white crossed region corresponds to the intersection of all constraints. EOS fulfilling all constraints are indicated by a blue symbol, a pentagon for the unique Skyrme model and a star for the three RMF ones.}
\label{L-J}
\end{figure}
%%%%%%%%%%%%%%%%%%%%%%%%%%%%%%%%%%%%%%%%%%%%%%%%%%%%%%%%%%%%%%%%%%%%%%%%%%%%%%%%%%%%%%%%%%

So far, two constraints were imposed on the various EOS discussed in Section~\ref{sec:unified}: the causality and ability to reach the $2\;M_\odot$ mass limit. There
are, however, several nuclear constraints that 
have been obtained from experiment or microscopic calculations during the last decade and which
 set much stronger conditions on the models.  In this subsection we impose in addition
  the following set of constraints reviewed in \cite{tsang12,lattimer13,lattimer14,dutra14}:
\begin{enumerate}
\item on the pressure for pure neutron matter from the calculations of
  \cite{H13} and of \cite{G12} (see Fig.~\ref{pn0:sk} and \ref{pn0:rmf}),
\item on the incompressibility of infinite nuclear matter at saturation $K=230\pm 40$ MeV \cite{khan2012},
\item constraints in the $J-L$ plane as compiled in \cite{lattimer14,tsang12} and plotted in Fig.~\ref{L-J}: 
\begin{enumerate}
\item from neutron skin thickness of $^{208}$Pb \cite{nskin},
\item from heavy ion collisions (HIC) \cite{HIC},
\item from electric dipole polarizability $\alpha_{\rm D}$ \cite{pola,lattimer14}, 
\item from giant dipole resonance (GDR) of $^{208}$Pb \cite{GDR}, 
\item from measured nuclear masses \cite{masses}, 
\item from isobaric analog states (IAS) \cite{IAS}.  
\end{enumerate} 
\end{enumerate} 
In Table~\ref{tab:constraints}, all our models are confronted with this
set of constraints: Y or N indicate whether the
constraint is satisfied or not, respectively. For the neutron pressure from
microscopic calculations we have also considered a less restrictive
constraint increasing by 10\% the uncertainty interval. 
The constraint
on the incompressibility is taken from  \cite{khan2012}. However, in  \cite{khan2013}
it was discussed that the uncertainty in the incompressibility is
related to the lack of knowledge of the skewness. Taking both Skyrme
interactions and RMF models the uncertainty in the skewness is larger
than $\pm 400$ MeV, so that the uncertainty of 17\% (corresponding to
40 MeV) obtained for the incompressibility in  \cite{khan2012} may be
underestimated taking into account that in their analysis only three
RMF models were considered. We, therefore, relax this constraint and
consider that NL3, NL3$\omega\rho$ and SK272 also satisfy the
constraint corresponding to increasing the uncertainty from 17\% to
18\%. This is indicated by the $^*$ symbol in Table~\ref{tab:constraints}.

\begin{table*}
\begin{ruledtabular}
\center\begin{tabular}{l|c|c|c|c|c|c|c|c|c|c|c}
Model & Hebeler & Hebeler$+10\%$ & Gandolfi & Gandolfi$+10\%$ & $K$ & Neutron skin & HIC & $\alpha_{\rm D}$ & GDR & masses & IAS\\
\hline
NL3     &N $     0.050 -     0.056$ &N $     0.050 -     0.061$ &N $     0.050 -     0.057$ &N $     0.050 -     0.062$ & Y*& N & N & N & N & N & N\\
{\bf NL3$\omega\rho$}   &N $     0.050 -     0.136$ &N $     0.050 -     0.155$ &N $     0.050 -     0.139$ &Y $     0.050 -     0.155$ & Y* & Y & Y & Y & Y & Y & Y\\ 
{\bf DDME2}   &Y $     0.050 -     0.155$ &Y $     0.050 -     0.155$ &Y $     0.050 -     0.155$ &Y $     0.050 -     0.155$ & Y & Y & Y & Y & Y & Y & Y\\ 
GM1     &N $     0.065 -     0.085$ &N $     0.061 -     0.092$ &N $     0.071 -     0.084$ &N $     0.066 -     0.091$ & N & N & Y & N & N & N & N\\ 
TM1     &N $     0.050 -     0.058$ &N $     0.050 -     0.063$ &N $     0.051 -     0.059$ &N $     0.050 -     0.063$ & N & N & N & N & N & N & N\\ 
DDHd    &N $     0.127 -     0.155$ &N $     0.120 -     0.155$ &N $     0.137 -     0.155$ &N $     0.129 -     0.155$ & Y & N & Y & N & N & N & N\\ 
{\bf DD2}     &N $     0.050 -     0.108$ &Y $     0.050 -     0.155$ &N $     0.050 -     0.087$ &Y $     0.050 -     0.155$ & Y & Y & Y & Y & Y & Y & Y\\ 
BSR2    &N $     0.065 -     0.116$ &N $     0.055 -     0.132$ &N $     0.078 -     0.113$ &N $     0.068 -     0.133$ & Y & Y & Y & Y & N & Y & N\\ 
BSR6    &N $     0.050 -     0.050$ &N $     0.050 -     0.053$ &N $     0.050 -     0.050$ &N $     0.050 -     0.055$ & Y & N & Y & Y & Y & N & Y\\ 
SKa     &N $     0.050 -     0.074$ &N $     0.050 -     0.089$ &N $     0.056 -     0.074$ &N $     0.050 -     0.085$ & Y & N & Y & Y & N & Y & Y\\ 
SKb     &N $     0.101 -     0.155$ &N $     0.094 -     0.155$ &N $     0.113 -     0.155$ &N $     0.104 -     0.155$ & Y & Y & N & N & N & N & N\\ 
SkI2    &N $     0.057 -     0.074$ &N $     0.054 -     0.081$ &N $     0.063 -     0.073$ &N $     0.058 -     0.080$ & Y & N & N & N & N & N & N\\ 
SkI3    &N $     0.058 -     0.080$ &N $     0.054 -     0.089$ &N $     0.065 -     0.079$ &N $     0.059 -     0.087$ & Y & N & Y & N & N & Y & N\\ 
SkI4    &N $     0.090 -     0.155$ &N $     0.082 -     0.155$ &N $     0.104 -     0.155$ &N $     0.093 -     0.155$ & Y & Y & Y & N & N & N & N\\ 
SkI5    &N $     0.055 -     0.066$ &N $     0.052 -     0.071$ &N $     0.059 -     0.067$ &N $     0.055 -     0.071$ & Y & N & N & N & N & N & N\\ 
SkI6    &N $     0.085 -     0.155$ &N $     0.076 -     0.155$ &N $     0.099 -     0.155$ &N $     0.089 -     0.155$ & Y & Y & Y & Y & N & Y & N\\ 
SLY2    &Y $     0.050 -     0.155$ &Y $     0.050 -     0.155$ &N $     0.087 -     0.155$ &Y $     0.050 -     0.155$ & Y & Y & N & N & Y & Y & Y\\ 
SLY230a &N $     0.104 -     0.155$ &Y $     0.050 -     0.155$ &N $     0.138 -     0.155$ &N $     0.113 -     0.155$ & Y & Y & N & N & N & Y & Y\\ 
SLY4    &Y $     0.050 -     0.155$ &Y $     0.050 -     0.155$ &N $     0.102 -     0.155$ &N $     0.067 -     0.155$ & Y & Y & N & N & Y & Y & Y\\
{\bf SLY9}    &N $     0.050 -     0.153$ &Y $     0.050 -     0.155$ &N $     0.070 -     0.155$ &Y $     0.050 -     0.155$ & Y & Y & Y & Y & Y & Y & Y\\ 
SkMP    &N $     0.064 -     0.101$ &N $     0.058 -     0.121$ &N $     0.072 -     0.097$ &N $     0.065 -     0.118$ & Y & Y & Y & N & N & N & N\\ 
SKOp    &N $     0.053 -     0.094$ &N $     0.050 -     0.128$ &N $     0.062 -     0.089$ &N $     0.054 -     0.129$ & Y & N & Y & Y & N & Y & N\\ 
KDE0v1  &Y $     0.050 -     0.155$ &Y $     0.050 -     0.155$ &Y $     0.050 -     0.155$ &Y $     0.050 -     0.155$ & Y & Y & N & N & N & N & N\\ 
SK255   &N $     0.050 -     0.050$ &N $     0.050 -     0.057$ &N $     0.050 -     0.052$ &N $     0.050 -     0.058$ & Y & N & N & N & N & N & N\\ 
SK272   &N $     0.050 -     0.050$ &N $     0.050 -     0.057$ &N $     0.050 -     0.052$ &N $     0.050 -     0.059$ & Y* & N & N & N & N & N & N\\ 
Rs      &N $     0.061 -     0.082$ &N $     0.056 -     0.091$ &N $     0.066 -     0.080$ &N $     0.062 -     0.089$ & Y & N & N & N & N & N & N\\ 
BSk20   &N $     0.098 -     0.155$ &Y $     0.050 -     0.155$ &N $     0.141 -     0.155$ &N $     0.110 -     0.155$ & Y & Y & N & N & Y & Y & Y\\ 
BSk21   &N $     0.109 -     0.155$ &N $     0.096 -     0.155$ &N $     0.126 -     0.155$ &N $     0.113 -     0.155$ & Y & Y & Y & Y & Y & Y & N\\ 
BSk22   &N $     0.068 -     0.124$ &N $     0.057 -     0.142$ &N $     0.082 -     0.123$ &N $     0.071 -     0.148$ & Y & N & Y & Y & N & Y & N\\ 
BSk23   &N $     0.083 -     0.155$ &N $     0.071 -     0.155$ &N $     0.099 -     0.155$ &N $     0.087 -     0.155$ & Y & Y & Y & Y & N & Y & N\\ 
BSk24   &N $     0.109 -     0.155$ &N $     0.096 -     0.155$ &N $     0.126 -     0.155$ &N $     0.113 -     0.155$ & Y & Y & Y & Y & Y & Y & N\\ 
BSk25   &N $     0.141 -     0.155$ &N $     0.131 -     0.155$ &N $     0.155 -     0.155$ &N $     0.147 -     0.155$ & Y & N & N & Y & Y & Y & N\\ 
BSk26   &N $     0.096 -     0.155$ &Y $     0.050 -     0.155$ &N $     0.138 -     0.155$ &N $     0.108 -     0.155$ & Y & Y & N & N & Y & Y & Y\\
\end{tabular}
\caption{Confrontation of the EOS with the various constraints. For
  each model the symbols Y and N indicate whether a given constraint
  is fulfilled or not, respectively. For the constraints from
  \cite{H13} and \cite{G12}, the interval of density over which the
  constraint is fulfilled is also given (in fm$^{-3}$). The $^*$ symbol indicates models for which the uncertainty on $K$ has been increased from 17\% to
18\%; see text for details.}
\label{tab:constraints}
\end{ruledtabular}
\end{table*}

Only one model satisfies all the constraints: DDME2. Increasing the
uncertainty interval of the neutron pressure from the calculations of
  \cite{H13} and \cite{G12}, three  more  models can be selected:
  DD2, NL3$\omega\rho$, SLy9. Their properties are summarized in Table~\ref{table:final}
  and 
in   Fig.~\ref{rad} we present the three selected RMF models with a
blue star and the  one Skyrme interaction with a blue pentagon.

\begin{center}
\begin{table*}
\begin{ruledtabular}
\begin{tabular}{lllllllllccl}
Model & $n_{\rm s}$  & $E_{\rm s}$ & $K$      & $J$   & $L$   & 
 $K_{\rm sym}$ & $M_{\rm max}^{\rm noY}$  &$n_{\rm DU} $& $M_{\rm DU}$\\
    & ${\rm fm^{-3}}$  & ${\rm MeV}$ & ${\rm MeV}$
   & ${\rm MeV}$   & ${\rm MeV}$   & ${\rm MeV}$ & 
  $M_\odot$  & ${\rm fm^{-3}}$ & $M_\odot$\\\hline
Sly9                & 0.151 & -15.8 & 229.8  & 32.0 &  54.9 & -81.42 & 2.16 & 0.56 &1.72 \\
NL3$_{\omega \rho}$ & 0.148 & -16.2 & 271.6  & 31.7 &  55.5 & -7.6 & 2.75 & 0.50 & 2.55\\
DDME2               & 0.152 & -16.1 & 250.9  & 32.3 &  51.2 & -87.1  & 2.48 &0.54 & 2.29\\
DD2                 & 0.149 & -16.0 & 242.6  & 31.7 &  55.0 & -93.2  & 2.42 & 0.54 & 2.18\\
\end{tabular}
\caption{Nuclear and astrophysical properties of models fulfilling all constraints.}
\label{table:final}
\end{ruledtabular}
\end{table*}
\end{center}

\begin{figure}
\resizebox{\hsize}{!}{\includegraphics{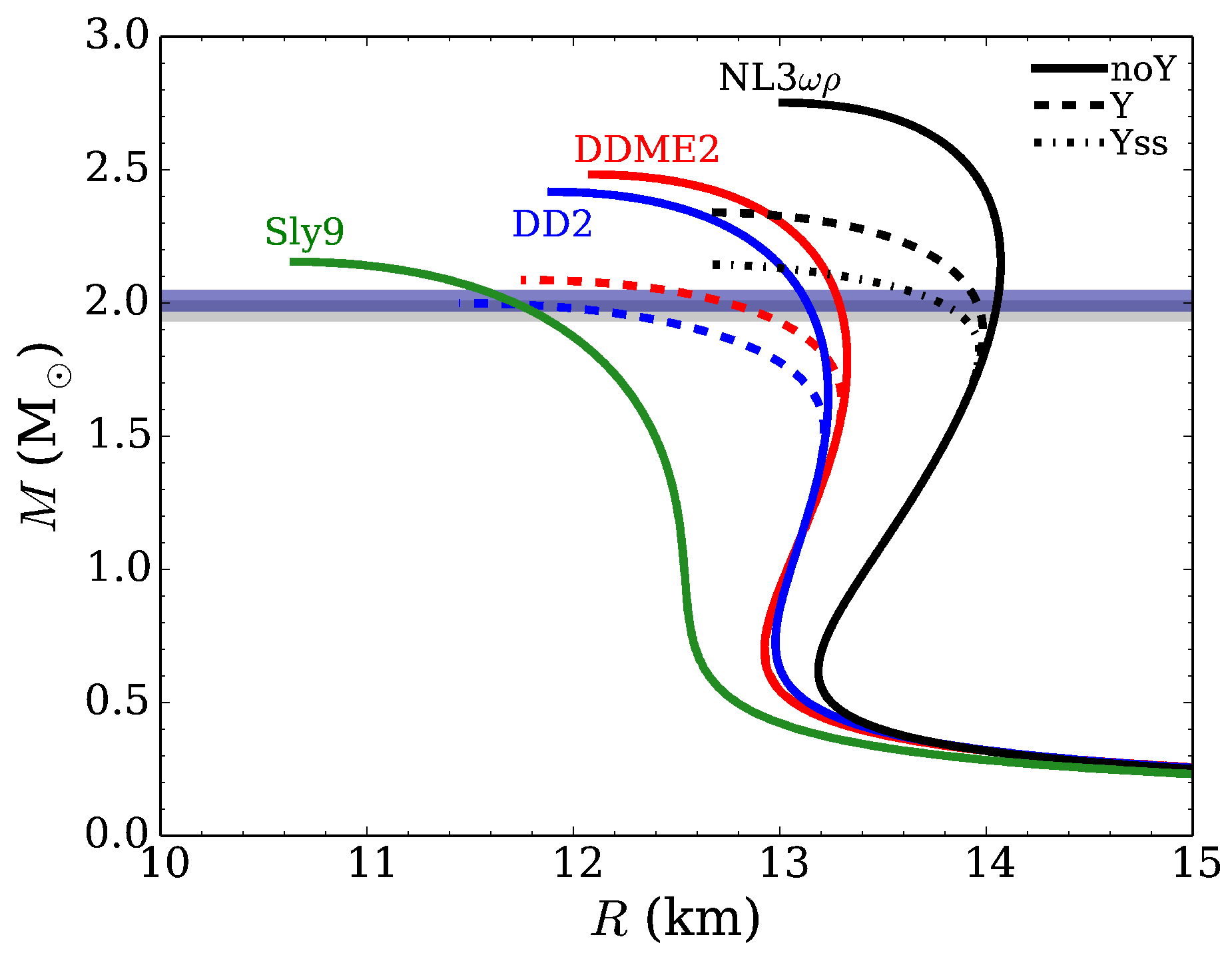}}
\caption{Mass-radius relations for the models fulfilling all nuclear constraints.}
\label{MRselect}
\end{figure}

In Fig.~\ref{MRselect} the mass-radius curves of the selected models
are shown for EOS of purely nucleonic and hyperonic matter (if available). 
Although they all have a very similar $L$: three of them have
$L\sim 55$ MeV and for the last one $L=51.2$ MeV, 
the radius uncertainty of a $1.4\;M_\odot$ star spanned by these models
is  $\Delta R_{1.4}=1.30$ km defined by the difference between $R_{1.4}(\mbox{SLy9})=12.45$ km and $R_{1.4}(\mbox{NL3}\omega\rho)=13.75$ km. This uncertainty reduces to 0.88 km for
$1.0\;M_\odot$ stars and increases to 2.34 km for $2.0\;M_\odot$ stars.
This radius interval is $\sim~1/3$ of the one that was obtained in Section~\ref{sec:unified} for $1.0\;M_\odot$ stars and
  $\sim~1/2$ for  $2.0\, M_\odot$ stars.
The fact that the range of possible radii is larger for the more massive
stars reflects the fact
that the high density range of the EOS is less well constrained than the
one close to and below saturation
density.

One property that is very different for all the four models represented by a blue mark
is the incompressibility $K$, see Fig.~\ref{rad},
bottom panels and Table~\ref{table:final}. The radius of the star is also to some extent
correlated with the incompressibility but, as discussed above, the uncertainty of the linear correlation is too large to provide a further 
constraint.

 The same set of experimental constraints employed in this work has been previously
used in \cite{lattimer14} with the same aim of addressing the relation between
uncertainties on the nuclear EOS and NS radii.
Using analytic equations between NS radii and 
pressure of beta-equilibrated matter, which in turn depends on $L$ and $K$,
the interval of $12.1\pm 1.1$ km was proposed, within 90\% confidence, for the radius 
of $1.4\;M_{\odot}$.

Finally, let us focus on the crust properties of the selected RMF models,
which have been plotted together in the upper panel of Fig.~\ref{crust}.
The dispersion observed in the crust thickness becomes
narrower, corresponding to $\sim 250$ m which represents $\sim30\%$ of the range obtained for all RMF models.

We can conclude that the present knowledge of $L$ and $K$ 
allows determining the NS radius within $1\;{\rm km}-2\;{\rm km}$. This residual uncertainty appears to be essentially due to  the lacking information on higher order terms, meaning that an increasing precision 
in the constraints for $L$ and $K$ is going to improve this
prediction only marginally. 

This underlines the importance of independent constraints.
One possibility would be to get information on higher order coefficients
(skewness, symmetry incompressibility, ..) from high density laboratory observables.

%L-J VS TSANG ET AL, 2012 %%%%%%%%%%%%%%%%%%%%%%%%%%%%%%%%%%%%%%%%%%%%%%%%%%%%%%%%%%%%%%%%%%%%%%%%%%%%%%%%%%%%%%%%%
%%%%%%%%%%%%%%%%%%%%%%%%%%%%%%%%%%%%%%%%%%%%%%%%%%%%%%%%%%%%%%%%%%%%%%%%%%%%%%%%%%%%%%%%%%

\subsection{ DUrca processes}
An interesting way to constrain the EOS could be to exploit independent astrophysical data, 
notably from the luminosity curves of the accreting NS and their interplay with the possible 
occurrence of the  DUrca process. This connection is explained in the following.

\begin{figure}
\resizebox{\hsize}{!}{\includegraphics{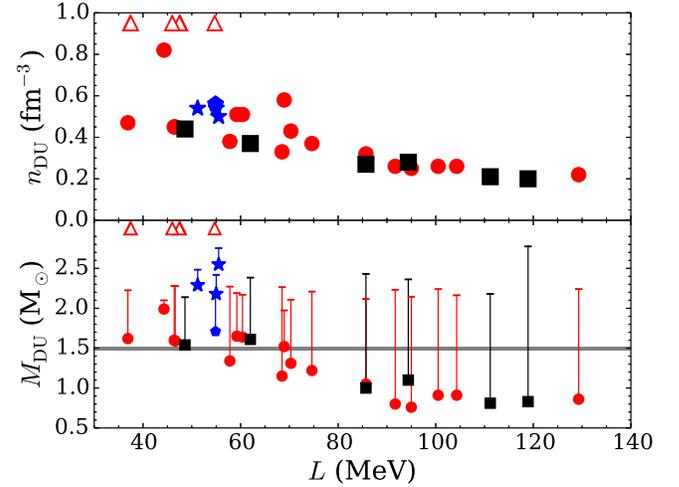}}
\caption{Density threshold $n_{\rm DU}$ (upper plot) and mass threshold $M_{\rm DU}$ (lower plot) for the nucleonic DUrca process to operate in purely nucleonic NS versus the slope of the symmetry energy $L$. The convention for colors and symbols is the same as in Fig.~\ref{rad}. Empty symbols on the upper x-axis indicate EOS for which $n_{\rm DU}$ is larger than the central density of the most-massive NS. In the lower plot, the error bars indicate the mass range over which the DUrca process operates.}
\label{nDUrcaL}
\end{figure}

After their birth in supernova explosions, NS are efficiently cooled down by neutrino emission during $\sim 10^5-10^6$ years (see \cite{HPY} and references therein). The simplest possible and most powerful neutrino process is the so-called nucleonic electron DUrca process \cite{LP91}:
\begin{equation}
n \rightarrow p + e^- + \bar{\nu}_e  \qquad \textrm{and} \qquad p+e^- \rightarrow n+\nu_e,
\label{eqn:DUrcanpe}
\end{equation}
which corresponds to the neutron $\beta$-decay followed by the electron capture on the proton.
Momentum conservation has to be satisfied for this process to operate which translates into the so-called triangle inequalities:
\begin{equation}
p_{{\rm F}n}\leq p_{{\rm F}p} + p_{{\rm F}e},
\end{equation}
where $p_{{\rm F}i}$ is the Fermi momentum of a species $i$.
Such inequalities impose a minimum proton fraction $Y_p^{\rm min}$ for the nucleonic DUrca process to occur \cite{klahn2006}: 
\begin{equation}
Y_{p}^{\rm min}=\frac{1}{1+\left(1+x_e^{1/3}\right)^{3}},
\end{equation}
with $x_e=n_e/\left(n_e+n_\mu\right)$. The absence of muons
corresponds to $x_e$=1 and $Y_{p}^{\rm min}=1/9$ while their
inclusion results in an increase of the value of $Y_{p}^{\rm min}$.
This minimum proton fraction translates into a threshold density $n_{\rm DU}$ and mass $M_{\rm DU}$ above which the DUrca process is active. A process similar to the one in Eq.~(\ref{eqn:DUrcanpe}) but involving muons instead of electrons may also operate; its threshold density is then slightly larger than for the electron DUrca process.

Mass and density threshold for the nucleonic DUrca process in purely nucleonic cores are given in Tables~\ref{tab:rmf} and \ref{table:NNparam} for RMF and Skyrme models, respectively. For some EOS, the density threshold above which the DUrca process operates is larger than the central density of the maximum-mass configuration. In other words, for such EOS the DUrca process is turned off for all possible masses and does not operate for any NS configuration.
In Fig.~\ref{nDUrcaL} the density threshold above which the DUrca
process operates in a purely nucleonic NS and the mass of the star with the corresponding central density are 
plotted against the slope of the symmetry energy. 
It reveals the possible existence of two distinct regions defined by a threshold on $L$: $L^{\rm DU}\simeq 70$ MeV.
Every non-hyperonic EOS with $L\gtrsim L^{\rm DU}$ has the DUrca process operating
in NS above a mass $M<1.5 M_\odot$. This is not the case for EOS
with $L\lesssim L^{\rm DU}$  as for some EOS the DUrca does not
operate at any NS mass and, for other EOS, it does for masses either above $2 M_\odot$ or
close to $1.5 M_\odot$. 
 In what regards the EOS that fulfill all constraints in Table~\ref{table:final} two patterns are to be noted.
For SLy9 the DUrca process is possible for masses larger than $1.72\, M_\odot$ while the three RMF models 
are characterized by DUrca thresholds above $2\; M_\odot$.

On the one hand, the thermal state of SAX J1808.4$-$3658, the coldest
observed transiently accreting NS, can be explained as shown in
\cite{BY15} by a very large neutrino emission in the core of NS that
only the very efficient DUrca process can explain.  Interestingly the region where all
nuclear constraints in the $L-J$ plane overlap, as plotted in Fig.~\ref{L-J},  
corresponds to values of $L$  that are strictly smaller
than the same $L^{\rm DU}$ below which the DUrca process does not
necessarily operate in massive purely nucleonic NS. Therefore reconciling the
nuclear constraints on $L$ and $J$ with the astrophysical one that the
DUrca process operates in NS might be challenging, as shown by the
fact that out of four EOS fulfilling our set of constraints only one
(SLy9) allows for the  DUrca process below $2 M_\odot$. In other 
words astrophysical observations of NS with a low luminosity might
 constrain the value of $L$ and consequently the radius as discussed
  in Section~\ref{sec:rad}.
On the other hand, population synthesis of isolated NS imposes that
the DUrca process does not occur in NS with masses $1.0 - 1.5\,M_\odot$
 \cite{PG06}, constraint that only SLy9  fulfill, unless a  strong proton
  superfluidity occurs in the core of low mass stars \cite{BY15}.\\

DUrca processes similar to the nucleonic ones can also occur in hyperonic NS \cite{PP92}. 
Examples relevant for our EOS are 
\begin{eqnarray}
\Lambda \rightarrow p + l + \bar{\nu}_l  \qquad &\textrm{and }&~~  p+l \rightarrow \Lambda+ \nu_l, \nonumber \\
\Xi^- \rightarrow \Lambda + l + \bar{\nu}_l  \qquad &\textrm{and }&~~  \Lambda + l \rightarrow \Xi^- + \nu_l,~ \nonumber \\ 
\Xi^- \rightarrow \Xi^0 + l + \bar{\nu}_l  \qquad &\textrm{and }&~~  \Xi^0 + l \rightarrow \Xi^- + \nu_l.~~
\label{eqn:DUrcaY}
\end{eqnarray}
The DUrca process involving a given hyperon turns on at a density very close to the onset density 
of this specific hyperon (with the condition that all other species involved in the process are also present). 
These density thresholds, or equivalently mass thresholds, for our RMF hyperonic EOS are given in Table~\ref{rmf:compo} together
with the same quantity for the nucleonic DUrca process. The hyperonic DUrca processes have weaker emissivities than their nucleonic counterparts \cite{PP92} but for some EOS they actually turn on at densities lower than the threshold for the nucleonic process. For the hyperonic version of our selected EOS: DD2, DDME2 and
NL3${\omega\rho}$, the mass threshold is 1.32, 1.46 and $1.59\, M_\odot$,  respectively. Nevertheless, a systematic study of the dependence of these thresholds on the poorly constrained hyperon properties and on the nuclear parameters (eg. $L$) is beyond the scope of this paper and will be the subject of future works.
It should be reminded that in the presence of hyperons the nucleonic DUrca process occurs at lower densities than expected in purely nucleonic stars due to the smaller neutron fraction. This could be an indication that  it is necessary to take into account hyperons in order to  reconcile an efficient DUrca process and an $L$ restricted to the interval allowed by terrestrial experiments.

\section{Conclusion}

 The present study has two  main objectives: (i) to illustrate the uncertainty that arises in the star radius  determination when  a non-unified EOS is used   for the integration of the TOV equations  and  (ii)  to quantify the same uncertainty taking a set of causal unified EOS that are consistent with the $2\,M_\odot$ maximum-mass limit, with or without considering an extra set of  constraints.

The unified EOS that are presented have been chosen among the nuclear RMF models and Skyrme interactions. In the latter case we have only considered models with causal EOS for densities 
 at least as high as the central density of a $2\,M_\odot$ star.
We have also considered EOS with  hyperonic degrees of freedom for all the  chosen RMF models. Except for DDH$\delta$ and TM1, all the other hyperonic EOS could still describe a $2\,M_\odot$ star when obtained using SU(6) symmetry to fix the vector meson coupling, experimental results to fix the scalar meson couplings, and, considering the mesons with hidden strangeness, including  the $\phi$ meson and excluding the $\sigma^*$ meson.

The unified EOS were built using different approaches for the RMF and Skyrme models.
For the  RMF EOS we take  the same  outer crust EOS \cite{ruester06} for all the models,  the inner crust is obtained within a Thomas Fermi calculation performed allowing for
non-spherical clusters according to \cite{GP0,GP}, and  the core is described by  the homogeneous matter EOS. The EOS are not completely unified due to the outer crust EOS, since this EOS is mainly fixed by experimental measurements; the effect of this approximation is however very small.
 Considering the non-relativistic unified EOS: at  low density 
 the nucleus $A$ and $Z$ numbers, as well as the volume $V_{\rm WS}$ of the Wigner-Seitz cell and the density of the free neutron component after drip are  variationally determined. The free neutrons are described with the  same functional used for the calculation of the core EOS. Concerning the nucleus, it is modeled with a compressible liquid-drop model with parameters fitted from Hartree-Fock calculations employing 
 the same Skyrme functional.

It was shown that for the non-unified EOS the crust-core matching may  quite strongly affect 
the radius  and crust thickness of the less massive stars. For our examples, depending on the matching procedure the difference in the radius and in crust thickness for a $1.0\,M_\odot$ star can be as large as  $\sim 1$ km and  $\sim 0.5$ km, respectively. This corresponds to relative differences as large as $\sim 4$\% for the radius and 30\% for the crust thickness. The largest uncertainties occur when the density dependence  of the symmetry energy, i.e. different slopes $L$,  of the crust and the core EOS are very different. This uncertainty may be minimized if  EOS for the crust and the core  with similar saturation properties are considered, when a unified EOS is not available.

Taking the initial set of EOS we have shown that the spanned  range of radii is $\sim 3$ km and $\sim 4$ km wide for $1.0\,M_\odot$ and $2.0\,M_\odot$ respectively. Imposing further constraints from experiment and theoretical calculations of neutron matter these intervals for radii are reduced respectively, to $\sim 1$ and 2 km.   Although smaller, this uncertainty is still large and reflects mostly our ignorance on the high density EOS, 
or equivalently on the higher order terms of the density expansion of the energy functional.
Additional uncertainties arise when the hyperon degrees of freedom are available. If hyperons are considered it is still possible to get $2\,M_\odot$ stars, meaning that they cannot be simply neglected. Stars with a mass $\gtrsim 1.5\, M_\odot$ typically contain hyperons in their core, and their presence is felt for the larger masses giving rise to a reduction of the star radius, and increasing uncertainties due to the largely unknown hyperon couplings.

Taking the whole set of models discussed in the Section~\ref{sec:unified} we have confirmed the existence of a linear correlation between the symmetry energy slope and the radius. 
This correlation is stronger for the less massive stars when the central densities of the stars are below 2.5$\rho_0$, a conclusion first drawn  in \cite{carriere03}. When larger masses are considered the spread of data increases, reflecting the lack of constraint to be imposed on the high density segment of the EOS. Considering the correlation between the incompressibility and the radius, the spread of data is independent of the mass of the star and 
prevents from extracting a clear correlation.
These results imply that further tighter constraints on $L$ and $K$ are not expected to improve the radius uncertainty in an important way. 

A very promising avenue is given by the potential constraint imposed by the necessity of DUrca processes to operate in NS in order to explain the observations of thermal states of some of them as shown in \cite{BY15}.
Indeed, the restricted $L$ interval compatible with present terrestrial constraints is close to the threshold for DUrca process in nucleonic stars. This means that only a limited number of functionals can at the same time fulfill the $L$ constraint and allow DUrca processes in NS. However, it was also shown that if  hyperons  are included, the  nucleonic DUrca process  is shifted to lower densities possibly  allowing to reconcile an efficient DUrca process and
an $L$ restricted to the interval allowed by terrestrial experiments.
This effect could be an indication of the presence of hyperons in the
interior of a neutron star. A further constraint might be obtained if the mass of a NS with a low luminosity is measured.

 Having shown the importance of using an unified EOS, all the studied EOS are accessible in the supplementary material section and on the CompOSE database.
\begin{acknowledgments}
The authors thank Micaela Oertel for the helpful discussions and assistance with the CompOSE database and Helena Pais for supplying some RMF inner crust EOS. This work has been partially supported by New-Compstar, COST Action MP1304, and by the Polish NCN grant no. 2014/13/B/ST9/02621. 
\end{acknowledgments}


\begin{thebibliography}{}{
\bibitem{demorest} P. B. Demorest, T. Pennucci, S. M. Ransom, M.S. E. Roberts, J. W. T. Hessels, Nature 467, 1081 (2010).
\bibitem{antoniadis} J. Antoniadis {\em et al.}, Science 340, 6131 (2013).
\bibitem{P14} A.~Y. Potekhin, Physics Uspekhi 57, 735 (2014).
\bibitem{steinerEPJA2015} A. W. Steiner,  J. M. Lattimer, E.F. Brown, Eur. Phys. J. A arXiv:1510.07515 (in press). 
\bibitem{fortin2015} M. Fortin, J.~L. Zdunik, P. Haensel and M. Bejger, \aap {\bf 576}, A68 (2015).
\bibitem{chen2015} Wei-Chia Chen, J. Piekarewicz, Phys. Rev. Lett. {\bf 115}, 161101 (2015). 
\bibitem{nicer} K.~C. Gendreau,  Z. Arzoumanian and T. Okajima, \procspie, 8443 (2012).
\bibitem{athena+} C. Motch, J. Wilms, D. Barret {\em et al.} , arXiv:1306.2334 (2013).
\bibitem{loft} M. Feroci, J.~W. den Herder, E. Bozzo {\em et al.} \procspie 8443 (2012).
\bibitem{glendenning99} N.~K. Glendenning, {\it Compact stars : nuclear physics, particle physics, and general relativity} (Springer, New York, 2000) .
\bibitem{bps} G. Baym, C. Pethick and P. Sutherland, ApJ {\bf 170}, 299 (1971).
\bibitem{Read} J.~S. Read, B.~D. Lackey, B.~J. Owen and J.~L. Friedman, Phys. Rev. {\bf D79}, 124032 (2009).
\bibitem{DHb} F. Douchin and P. Haensel, A\&A {\bf 380}, 151 (2001).
\bibitem{DHa} E. Chabanat, P. Bonche, P. Haensel, J. Meyer and R. Schaeffer, Nucl. Phys. {\bf A 635}, 231 (1998).
\bibitem{VN} J.~W. Negele and D. Vautherin, Nucl. Phys. {\bf A 207}, 298 (1973).
\bibitem{piekarewicz2014} J. Piekarewicz, F.J. Fattoyev, C.J. Horowitz, Phys. Rev. C {\bf 90}, 015803 (2014).
\bibitem{BHF} H.-J. Schulze and T. Rijken, Phys. Rev. C {\bf 84}, 035801 (2011);
Eur. Phys. J. A  {\bf 52}, 21 (2016).
\bibitem{pederiva} D. Lonardoni, A. Lovato, S. Gandolfi, F. Pederiva, Phys. Rev. Lett. {\bf 114}, 092301 (2015).
\bibitem{BY15} M.~V. Beznogov and D.~G. Yakovlev, \mnras ~ {\bf 447}, 1598 (2015).
\bibitem{tov} J. R. Oppenheimer and G. M. Volkoff, Phys. Rev. {\bf 55}, 374 (1939); R. C. Tolman, {\it ibid.} {\bf 55}, 364 (1939).
\bibitem{hp} P. Haensel and B. Pichon, A\&A 283, 313 (1994).
\bibitem{ruester06} S.~B. R{\"u}ster, M. Hempel and J. Schaffner-Bielich, Phys. Rev. C {\bf 73}, 035804 (2006).
\bibitem{ravenhall83} D.~G. Ravenhall, C.~J. Pethick and J.~R. Wilson, Phys. Rev. Lett. {\bf 50}, 2066 (1983).
\bibitem{horowitz04} C.~J. Horowitz, M.~A. P{\'e}rez-Garc{\'{\i}}a and J. Piekarewicz,  Phys. Rev. C  {\bf 69}, 045804 (2004).
\bibitem{horowitz05} C.~J. Horowitz, M.~A. P{\'e}rez-Garc{\'{\i}}a, D.~K. Berry and J. Piekarewicz, Phys. Rev. C {\bf 72}, 035801 (2005).
\bibitem{maruyama05} T. Maruyama, T. Tatsumi, D. N. Voskresensky, T. Tanigawa and S. Chiba, Phys. Rev. C {\bf 72}, 015802 (2005).
\bibitem{sonoda} H. Sonoda, G. Watanabe, K. Sato, K. Yasuoka, and T. Ebisuzaki, Phys. Rev. C {\bf 77}, 035806 (2008).
\bibitem{sonoda2} G. Watanabe, H. Sonoda, T. Maruyama  {\em et al.}, Phys. Rev. Lett. {\bf 103}, 121101 (2009).
\bibitem{avancini08} S. S. Avancini, D. P. Menezes, M. D. Alloy, J. R. Marinelli, M. M. W. de Moraes and 
C. Provid\^{e}ncia, Phys. Rev. C {\bf 78}, 015802 (2008); 
S. S. Avancini, L. Brito, J. R. Marinelli, D. P. Menezes, M. M. W. de Moraes, C. Provid\^{e}ncia, and A. M. Santos, 
Phys. Rev. C {\bf 79}, 035804 (2009); 
S. S. Avancini, S. Chiacchiera, D. P. Menezes, and C. Provid\^{e}ncia, Phys. Rev. C {\bf 82}, 055807 (2010); 
Phys. Rev. C {\bf 85}, 059904(E) (2012).
\bibitem{newton09}W. G. Newton and J. R. Stone, Phys. Rev. C 79,  055801 (2009).
\bibitem{YL14} Yasutake, N., {\L}astowiecki, R., Beni{\'c}, S., et al.\ 2014, \prc, 89, 065803 
\bibitem{bbp} G. Baym, H.~A. Bethe, and C.~J. Pethick, Nucl. Phys. {\bf A 175}, 225 (1971).
\bibitem{NV} J.~W. Negele, D.  Vautherin,  Nucl. Phys. {\bf A 207}, 298 (1973).
\bibitem{GP} F. Grill, H. Pais, C. Provid{\^e}ncia, I. Vida{\~n}a, and S.~S. Avancini, Phys. Rev. C {\bf 90}, 045803 (2014).
\bibitem{GM1} N.~K. Glendenning, and  S.~A. Moszkowski, Phys. Rev. Lett. {\bf 67}, 2414 (1991).
\bibitem{NL3} G.~A. Lalazissis, J. K{\"o}nig, and P. Ring, Phys. Rev. C {\bf 55}, 540 (1997).
\bibitem{NL3wra} C.~J. Horowitz, and J. Piekarewicz,  Phys. Rev. Lett. {\bf 86}, 5647 (2001).
\bibitem{SY07} Shternin, P.~S., Yakovlev, D.~G., Haensel, P., \& Potekhin, A.~Y.\ 2007, \mnras, 382, L43 
\bibitem{PR13} Page, D., \& Reddy, S.\ 2013, Physical Review Letters, 111, 241102 
\bibitem{AG12} Andersson, N., Glampedakis, K., Ho, W.~C.~G., \& Espinoza, C.~M.\ 2012, Physical Review Letters, 109, 241103 
\bibitem{Pi14} Piekarewicz, J., Fattoyev, F.~J., \& Horowitz, C.~J.\ 2014, \prc, 90, 015803 
\bibitem{GN11} Gearheart, M., Newton, W.~G., Hooker, J., \& Li, B.-A.\ 2011, \mnras, 418, 2343 
\bibitem{TM1} Y. Sugahara, and H. Toki, Nucl. Phys. {\bf A, 579}, 557 (1994).
\bibitem{stosa} H. Shen, H. Toki, K. Oyamatsu, and K. Sumiyoshi, Nucl. Phys. {\bf A 637}, 435 (1998).
\bibitem{stosb} H. Shen, H. Toki, K. Oyamatsu, and K. Sumiyoshi, Progress of Theoretical Physics {\bf 100}, 1013 (1998).
\bibitem{DK07} S.~K. Dhiman, R. Kumar, and B.~K. Agrawal, Phys. Rev. C {\bf 76}, 045801 (2007).
\bibitem{A10}  B.~K. Agrawal, Phys. Rev. C {\bf 81}, 034323 (2010).
\bibitem{DDME2} G.~A. Lalazissis, T. Nik{\v s}i{\'c}, D. Vretenar, and P. Ring, Phys. Rev. C {\bf 71}, 024312 (2005).
\bibitem{typel2009} S. Typel, G. R{\"o}pke, T. Kl{\"a}hn, D. Blaschke, and  H.~H. Wolter, Phys. Rev. C {\bf 81}, 015803 (2010).
\bibitem{G04} T. Gaitanos, M. Di Toro, S. Typel, {\em et al.}, Nucl. Phys. {\bf A 732}, 24 (2004).
\bibitem{BH14} S. Banik, M. Hempel, and D. Bandyopadhyay,  ApJS, {\bf 214}, 22 (2014).
\bibitem{GP0} F. Grill, C. Provid{\^e}ncia, and S.~S.  Avancini, Phys. Rev. C {\bf 85}, 055808 (2012).
\bibitem{MC13} T. Miyatsu, M.-K. Cheoun, and K. Saito, \prc, {\bf 88}, 015802 (2013).
\bibitem{SA12} A. Sulaksono, and B.~K. Agrawal, Nucl. Phys. {\bf A 895}, 44 (2012).
\bibitem{BH12} I. Bednarek, P. Haensel, J.~L. Zdunik,  M. Bejger, and R. Ma{\'n}ka,  \aap {\bf 543}, A157 (2012).
\bibitem{SBG00} J. Schaffner-Bielich, and A. Gal, , \prc {\bf 62}, 034311 (2000).
\bibitem{OP15} M. Oertel, C. Provid{\^e}ncia, F. Gulminelli, \& A.~R. Raduta, Journal of Physics G: Nuclear Physics, {\bf 42}, 075202 (2015).
\bibitem{T01} H. Takahashi, J.~K. Ahn, H. Akikawa, {\em et al.}, Phys. Rev. Lett. {\bf 87}, 212502 (2001).
\bibitem{oyamatsu07}  K. Oyamatsu and K. Iida, Phys. Rev. C 75, 015801 (2007)
\bibitem{SKab} H. S. Kohler, Nucl. Phys. {\bf A258}, 301 (1976).
\bibitem{SkI2-5} P.-G. Reinhard and H. Flocard, Nucl. Phys. {\bf A 584}, 467 (1995).
\bibitem{SkI6} W. Nazarewicz et al., Phys. Rev. C {\bf 53}, 740(1996) .
\bibitem{SLY2and9} E. Chabanat, Interactions effectives pour des conditions extremes d'isospin, Ph.D. thesis, University Claude Bernard Lyon-1, Lyon, France, 1995.
\bibitem{SLY230a} E. Chabanat et al., Nucl. Phys. {\bf A 627}, 710 (1997).
\bibitem{SLY4} E. Chabanat et al., Nucl. Phys. {\bf A 635} (1998) 231.
\bibitem{SkMP} L. Bennour et al., Phys. Rev. C {\bf 40}, 2834 (1989).
\bibitem{SKOp} P.-G. Reinhard et al., Phys. Rev. C {\bf 60}, 014316 (1999).
\bibitem{KDE0v} B. K. Agrawal, S. Shlomo and V. K. Au, Phys. Rev. C {\bf 72}, 014310 (2005).
\bibitem{SK255and272} B. K. Agrawal, S. Shlomo and V. Kim Au, Phys. Rev. C {\bf 68}, 031304 (2003).
\bibitem{Rs} J. Friedrich and P.-G. Reinhard, Phys. Rev. C {\bf 33}, 335 (1986).
\bibitem{BSk20-21} S. Goriely, N. Chamel, and J. M. Pearson, Phys. Rev. C 82, 035804 (2010).
\bibitem{BSk22-26} S. Goriely, N. Chamel, and J. M. Pearson, Phys. Rev. C 88, 024308 (2013).
\bibitem{WSpaper} F. Gulminelli, Ad.R. Raduta, Phys. Rev C {\bf 92}, 044313 (2015).
\bibitem{Danielewicz2009} P. Danielewicz and J. Lee,  Nucl. Phys. {\bf A818}, 36 (2009). 
\bibitem{ldm} W.  D.  Myers and W.  J.  Swiatecki, Nucl. Phys. {\bf A 336}, 267 (1980). 
\bibitem{Panagiota2013} P. Papakonstantinou, J. Margueron, F. Gulminelli, Ad. R. Raduta, 
                        Phys. Rev. C {\bf 88}, 045805 (2013). 
\bibitem{esym_paper} A. R. Raduta, F. Gulminelli and F. Aymard, Eur. Phys. J. {\bf A 50}, 24 (2014).
\bibitem{BBP} G. Baym, H. A. Bethe and C.Pethick,  Nucl. Phys. {\bf A175}, 225 (1971). 
\bibitem{douchin} F. Douchin, P. Haensel, Phys. Lett.  {\bf B 485}, 107 (2000). 
\bibitem{Aymard2014} F. Aymard, F. Gulminelli and J. Margueron, Phys. Rev. C {\bf 89}, 065807 (2014).
\bibitem{pasta_RMF} S. S. Avancini, D. P. Menezes, M. D. Alloy {\em et al.}, Phys. Rev.  {\bf C  78},  015802 (2008). 
\bibitem{lattimer07} J. M. Lattimer, Maddapa Prakash, Phys. Rept. {\bf 442}, 109 (2007).  
\bibitem{carriere03} J. Carriere, C.J. Horowitz, J. Piekarewicz,  Astrophys.J. 593, 463 (2003) .
\bibitem{brown00} B. A. Brown, Phys. Rev. Lett. {\bf 85}, 5296 (2000); 
                  S. Typel and B. A. Brown, Phys. Rev. C {\bf 64}, 027302 (2001); 
                  C. J. Horowitz and J. Piekarewicz, Phys. Rev. Lett. {\bf 86}, 5647 (2001); 
                  A. W. Steiner, M. Prakash, J. Lattimer and P. J. Ellis, Phys. Rep. {\bf 411}, 325 (2005).
\bibitem{vinas09} M. Centelles, X. Roca-Maza, X. Vinas, M. Warda, Phys. Rev. Lett. {\bf 102}, 122502 (2009). 
\bibitem{margueron15} J. Margueron, NewCompStar School 2016, \url{http://compstar15.nipne.ro/}
\bibitem{tsang12} M.~B. Tsang, J.~R. Stone, F. Camera, {\em et al.}, \prc {\bf 86}, 015803 (2012).
\bibitem{lattimer13}  J. M. Lattimer and Y. Lim, Astrophys. J. {\bf 771}, 51 (2013).
\bibitem{lattimer14} J. M. Lattimer and A. W. Steiner, EPJA {\bf 50}, 40  (2014). 
\bibitem{dutra14} M. Dutra, O. Louren{\c c}o, S.~S. Avancini, {\em et al.}, Phys. Rev. C {\bf 90}, 055203 (2014).
\bibitem{H13} K. Hebeler, J.~M. Lattimer, C.~J. Pethick, and A.  Schwenk, ApJ {\bf 773}, 11 (2013).
\bibitem{G12} S. Gandolfi, J. Carlson, and S. Reddy, Phys. Rev. C {\bf 85,} 032801 (2012).
\bibitem{khan2012} E. Khan, J. Margueron, I. Vidana,  Phys. Rev. Lett. {\bf 109}, 092501 (2012).
\bibitem{nskin} L.-W. Chen, C.~M. Ko, B.-A. Li, and J. Xu, \prc {\bf 82}, 024321 (2010).
\bibitem{HIC} M.~B. Tsang,  Y. Zhang, P. Danielewicz, {\em et al.},  Phys. Rev. Lett. {\bf 102}, 122701 (2009).
\bibitem{pola} X. Roca-Maza, M. Brenna,  G. Col{\`o}, {\em et al.}, \prc {\bf 88}, 024316 (2013).
\bibitem{GDR} L. Trippa, G. Col{\`o}, and E. Vigezzi, \prc {\bf 77}, 061304 (2008).
\bibitem{masses} M. Kortelainen, T. Lesinski, J. Mor{\'e}, {\em et al.} \prc {\bf 82}, 024313 (2010).
\bibitem{IAS} P. Danielewicz, and J. Lee, Nucl. Phys. {\bf A 922}, 1 (2014).
\bibitem{khan2013} E. Khan and J. Margueron, Phys. Rev C {\bf 88}, 034319 (2013).
\bibitem{HPY} P. Haensel, A.~Y. Potekhin, and D.~G. Yakovlev, {\it Neutron Stars 1. Equation of state and structure} (Springer, New York, 2007) .
\bibitem{LP91} J.~M. Lattimer, M. Prakash, C.~J. Pethick, and P. Haensel, Phys. Rev. Lett. {\bf 66}, 2701 (1991).
\bibitem{klahn2006} T. Klahn {\em et al.}, Phys. Rev. C {\bf 74}, 035802 (2006). 
\bibitem{PP92} M. Prakash, M. Prakash, J.~M.  Lattimer, and C.~J. Pethick, 
\apjl {\bf 390}, L77 (1992).
\bibitem{PG06} S. Popov, H. Grigorian, R. Turolla, and D. Blaschke, \aap, {\bf 448}, 327 (2006).



\bibliographystyle{unsrt}}


\end{thebibliography}
\end{document}